\documentclass[twocolumn, aps, prx, superscriptaddress,longbibliography]{revtex4-1}
\usepackage{amssymb,bbold,dsfont}
\usepackage{amsmath}
\usepackage{color}
\usepackage{graphicx}
\usepackage{marvosym}
\usepackage{notoccite}
\usepackage{comment}
\usepackage{pifont}
\usepackage{bbm}
\usepackage{graphicx,tabularx}
\usepackage{dcolumn}
\usepackage{bm}
\usepackage{xcolor}
\usepackage{marvosym}
\usepackage[breaklinks=true, colorlinks,citecolor=blue,linkcolor=blue,urlcolor=blue]{hyperref}
\usepackage{titlesec}
\newcommand{\be}{\begin{equation}}
\newcommand{\ee}{\end{equation}}

\begin{document}
	\title{Noisy information channel mediated prevention of the tragedy of the commons}

\author{Samrat Sohel Mondal}
\email{s5mondal@ucsd.edu }
\affiliation{
  Department of Ecology, Behavior, and Evolution, University of California San Diego, La Jolla, CA 92093
}

\author{Sagar Chakraborty}
\email{sagarc@iitk.ac.in}
\affiliation{
  Department of Physics, Indian Institute of Technology Kanpur, Uttar Pradesh 208016, India
}
\date{\today}

\begin{abstract}
	Synergy between evolutionary dynamics of cooperation and fluctuating state of shared resource being consumed by the cooperators is essential for averting the tragedy of the commons. Not only in humans, but also in the cognitively-limited organisms, this interplay between the resource and the cooperation is ubiquitously witnessed. The strategically interacting players engaged in such game-environment feedback scenarios naturally pick strategies based on their perception of the environmental state. Such perception invariably happens through some sensory information channels that the players are endowed with. The unfortunate reality is that any sensory channel must be noisy due to various factors; consequently, the perception of the environmental state becomes faulty rendering the players incapable of adopting the strategy that they otherwise would. Intriguingly, situation is not as bad as it sounds. Here we introduce the hitherto neglected information channel between players and the environment into the paradigm of stochastic evolutionary games with a view to bringing forward the counterintuitive possibility of emergence and sustenance of cooperation on account of the noise in the channel. Our primary study is in the simplest non-trivial setting of two-state stochastically fluctuating resource harnessed by a large unstructured population of cooperators and defectors adopting either memory-1 strategies or reactive strategies while engaged in repeated two-player interactions. The effect of noisy information channel in enhancing the cooperation in reactive-strategied population is unprecedented. We find that the propensity of cooperation in the population is inversely related to the mutual information (normalized by the channel capacity) of the corresponding information channel.
\end{abstract}
\maketitle

\section{Introduction}
Propaganda~\cite{Barfar2022} in politics and deceptive signals~\cite{Rowell2006} in ecosystems are among many other daily-life phenomena that shield the true nature of environment, leading to presumably suboptimal responses by the agents (or players) who strategically interact based on the state of the environment. For example, voters in an election held in a democracy would vote---based on their selfish reasons---either for corrupt candidates or an uncorrupt ones. However, the candidates' corruption-levels are known to the voters through the media which are somewhat controlled or influenced by the candidates; the voters' information about the candidates' corruption-levels  is at best noisy. Thus, the game which the voters are playing among themselves while electing the candidates is highly influenced by the noisy channel, viz., the media. Similarly, whether two predators would hunt together or go on separate ways to hunt alone might depend on how numerous the preys are or how difficult the preys are. The information about the preys' status has to be gained through various cues and signals present in the ecosystem. Be it due to deceptive signals sent by the preys or inability of the predators to understand the cues in the environment, the strategic decisions by the predators do get influenced. Once again noisy information channel is a paramount factor. 

Therefore, the influence of the noisy channel must not be ignored if one desires to understand a more complete picture of the feedback~\cite{weitz2016oscillating,lin2019prl,tilman2020nc,Wang2020,Barfuss2020,DasBairagya2021,Bairagya2023,SohelMondal2024} between what strategies players adopt and what the state of the environment is. A notable example of such a feedback loop is the tragedy of the commons~\cite{hardin1968commons, ostrom1999coping} (ToC), where mutually non-cooperating players degrade their environment~\cite{Ashcroft2014,Gokhale2016,Hauert2006,Tavoni2012}, thereby limiting their potential long-term benefits. While noise has a negative connotation associated with it, it is not hard to envisage that the information channel may be manipulated to influence the players' behaviour during interactions. Such a manipulation would be quite interesting and useful while tackling the ToC---the selfish overexploitation of public common goods.

The interplay between group behaviour and environment can be conceptualized as a stochastic game~\cite{Shapley1953,Hilbe2018,Kleshnina2023}, where individuals interact over multiple rounds. Their actions in one round can impact the environmental condition they face in the next, such as when populations aim to control  epidemics~\cite{Johnson2020,Chica2021,Abel2021}, manage natural resources~\cite{Samuelson1990,VanVugt2002,Cumming2017}, or mitigate climate change~\cite{Milfont2010,Tavoni2011,Vesely2020}. The interaction over multiple rounds finds its mathematization in the formalism of repeated games~\cite{SigmundBook} which serves as an evolutionary game theoretical paradigm in the exploration of reciprocity induced cooperation~\cite{Nowak2006}. Mutual cooperation would prevent the ToC.

Stochastic games have emerged as a valuable tool for understanding the evolution of cooperation in changing environments~\cite{Hilbe2018,Kleshnina2023}. One finds~\cite{Hilbe2018} that the dependence of the environmental state on previous interactions can increase the chances for cooperation compared to when  there are repeated interactions in an unchanging environment or when there are one-off interactions in a changing one. What is even more intriguing---in the context of the present paper---is that the fates of cooperation and environment depends crucially whether the players adopt their strategy in the presence of full information about the current state of the environment or in the complete absence of it~\cite{Kleshnina2023}. It begs the question: What if the information about the environmental state is incomplete? One can contemplate existence of a noisy information channel between the true state of the environment and individuals perceiving it.

In many real-world applications, any information about the state of the environment is, at best, incomplete. \textcolor{black}{Such uncertainties can, in turn, dramatically affect human behavior~\cite{Morton2011,Barrett2013,AbouChakra2018,Paarporn2018,Wang2024}.} Understanding the impact of incomplete information on decision-making has been a rich field of study in economics. Corresponding studies~\cite{Harsanyi1967,Levine1977,Bagh2019} suggest that the effect of information is often positive, even though there are situations in which it has adverse effects. Additionally, studies~\cite{Hansen2004,Barfuss2022} of partially observable stochastic games suggest that settings with incomplete information can benefit decision-makers.

The noisy information channel and its capacity, the cornerstone concept of the Shannon's communication theory, is a surprisingly overlooked factor in the investigation of feedback between sustenance of cooperation and prevention of the ToC. Here we include this aspect in the evolutionary game theoretic setup of stochastic games: Specifically, we investigate how the noisy processing of information of state by individuals shapes the evolution of cooperation in the population. The individuals are modelled to perceive the state of the shared environment incompletely owing to the imperfect sensory information channels and, thence, to condition their future actions on the perception about the current state. The result is exciting: Noisy channel can improve mutual cooperation.

Our convictions are best depicted in the mathematical setting of repeated games where the players in each round adopt memory-1 strategies or even simpler reactive (memory-$\frac{1}{2}$) strategies. In the former, a player adopts action in the subsequent round based on only the latest action-profile, while in the latter, a player's future action depends on the immediately preceding action of its opponent. For simplicity, it is assumed that at each round one-shot--two-player game is played among any two players in the unstructured large population. Furthermore, most importantly, we allow for the strategy of the players to be contingent on the {\it perceived} state of the environment. Using the concept of mutual information and channel capacity, we devise a measure of the {\it efficacy} of noisy information harnessing and find its role in characterizing the average information processing by the population in the sustenance of cooperation. Moreover,  to quantify the importance of noise, we measure how much the propensity of cooperation and the prevention of the ToC in a population enhances in the presence of noise in sensory channels.

\begin{figure}[t!]
	\centering
	\includegraphics[scale=0.51]{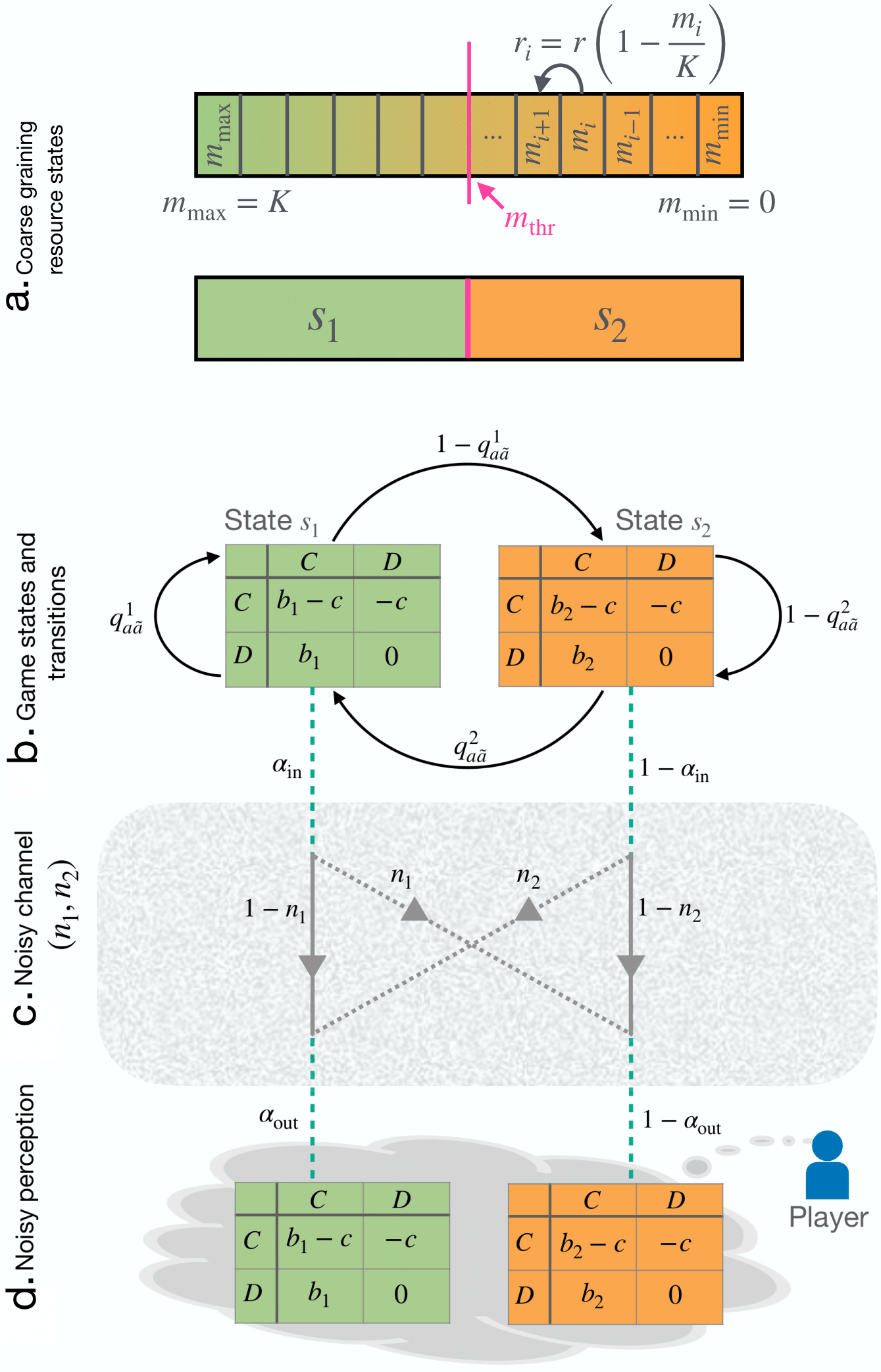}
	\caption{{\it Schematic illustrating the noisy-channel–mediated stochastic game:} {\color{black}{(a) depicts a  self-renewing, multi-state resource that is coarse-grained into a two-state representation on adopting a threshold $m_{\rm thr}$.}} (b) shows two possible states of the stochastic donation game. The green and the orange matrices, respectively, represent the more beneficial and the less beneficial states of the stochastic game. The transition from state $s_{i}$ to the other state $s_{1}$ happens with probability $q_{a\tilde{a}}^{i}$. The frequency of the true beneficial state is $\alpha_{\rm in}$. (c) depicts the noisy information channel $(n_{1},n_{2})$ through which players perceive state $s_{i}$ as $s_{\bar{i}}$ erroneously with probability $n_{i}$. (d) highlights the changed probability distribution, $(\alpha_{\rm out}, 1-\alpha_{\rm out})$, over the two states due to wrong perception of the true distribution, $(\alpha_{\rm in}, 1-\alpha_{\rm in})$. }
	\label{fig:cartoon}
\end{figure}

\section{Game Dynamics Through Noisy Channel}
We consider the simplest non-trivial setup of stochastic games, in which two players interact for infinitely many rounds. For concreteness and simplicity, we assume that the probability of every subsequent round---also known as discount factor~\cite{Hilbe2018}---is unity. \textcolor{black}{The effect of fractional discount factor is discussed in Appendix.~\ref{aShadowScheme}}. In each round, both players can  simultaneously be in one of the two possible states: $s_1$ and $s_2$. Players engage in one of two possible donation games depending on the current state. For both the games, each player has same action set containing two actions: to cooperate $(C)$ and to defect $(D)$. Here, action $C$ employed by a player entails that the player incurs a cost $c$ to render the opponent a benefit $b_i$ in the current state is $s_i$. Similarly, $D$ implies not providing the benefit while incurring no cost. One should note that while the cost remains $c$ irrespective of the state of the stochastic game, the value of the benefit provided to the opponent depends on the current state of the stochastic game. We assume the first state to be more beneficial to the opponent if the focal player pays the cost, so that $b_1\geq b_2 > c$; furthermore, we set $c=1$.

{\color{black}
Before proceeding further, a remark is in order: In real systems, usually, state of a resource is not binary but is often conventionally represented by a continuous variable ($m$, say); a large body of literature~\cite{Tsoularis2002,Mndez2015,tilman2020nc} models such resources using logistic growth with intrinsic rate $r>0$ (i.e., self-renewing resource) and carrying capacity $K>0$. To embed such a resource into the framework of a stochastic game, it is convenient to consider discretization of the continuous state, i.e.,  states $m_i$, where $i\in\{0,1,2,\cdots,M\}$, $m_0=m_{\min}=0$ and $m_M=m_{\max}=K$ (see Fig.~\ref{fig:cartoon}a). In the absence of  any action by players, the resource transitions from state $m_i$ to a more abundant state $m_{i+1}$ occurs at rate $r_i=r\left(1 - \frac{m_i}{K}\right)$. Obviously, $r_i<r_{i+1}$ $\forall i$ unless, of course, the resource is not self-renewing, i.e., $r=0$. In effect, in this paper, we have adopted a further simplification~\cite{Patra2025}: We coarse-grain the full set of states (i.e., $m_i$'s) into two exclusive and exhaustive categories. For this purpose, we choose a threshold, $m_{\rm thr}$, and assign all $m_i \ge m_{\rm thr}$ to a single beneficial resource state $s_1$, while the all remaining states form a single degraded resource state $s_2$. This simplified two-state representation (see Fig.~\ref{fig:cartoon}a) preserves the key qualitative features of the resource dynamics, provided that the stochastic game is chosen appropriately---a point we will discuss later in the text---while still keeping the analysis tractable.}

Since, in this study, we assume that players use the perception of the current state of the stochastic game and the latest action profile to choose action in the very next round, the memory-1 strategies of a player in the stochastic repeated game is an 8-tuple:
\begin{equation}\label{eq:strategyVector}
\bm{p} = (p_{CC}^1,p_{CD}^1,p_{DC}^1,p_{DD}^1;p_{CC}^2,p_{CD}^2,p_{DC}^2,p_{DD}^2).
\end{equation}
Here, $p_{a\tilde{a}}^j$ is the probability of cooperation by the player in the next round, given the latest action profile is $(a,\tilde{a})$ and the current perceived state of the environment is $s_{j}$. Note $a$ is the action of the focal player and $\tilde{a}$ is that of her opponent. A memory-1 strategy for which $p_{C\tilde{a}}^j = p_{D\tilde{a}}^j$ effectively makes it a reactive or  memory-$\tfrac{1}{2}$ strategy: A players with reactive strategy is seen to react only to the opponent's action. If all $p_{a\tilde{a}}^j\in\{0,1\}$, the corresponding strategy is termed pure.

However, on bringing this paper's core idea---viz., noisy information channel---into consideration, the aforementioned strategy is modified. To this end, supposing the current state to be $s_{j}$, the players perceive the state incorrectly with probability $n_{j}$ (and, of course, correctly with probability $1-n_{j}$).  Naturally, any probability $p_{a\tilde{a}}^j$ of cooperation by the player in the next round, should transform as
\begin{equation}\label{eq:channelTrans}
p_{a\tilde{a}}^j\to p_{a\tilde{a}}^{'j}=(1-n_j)p_{a\tilde{a}}^j+n_{j}p_{a\tilde{a}}^{\bar{j}},
\end{equation}
where $\bar{j}$ is defined as 1 or 2, respectively, for $j$ equal to 2 or 1. An information channel---henceforth, aptly indicated by $(n_1,n_2)$---with $n_{1}=n_{2}$ is said to be symmetric; otherwise, it is asymmetric. Henceforth, we drop the prime for notational convenience: the symbol $p_{a\tilde{a}}^j$ denotes $p_{a\tilde{a}}^{'j}$.

Another crucial ingredient worth specifying in the stochastic repeated game is how the states change over the rounds of play. \textcolor{black}{Here, we assume~\cite{Hilbe2018,Su2019,Wang2021,Kleshnina2023} that} the transitions between the states of the stochastic games are governed by the latest state $s_{i}$ and action profile $(a,\tilde{a})$.  Thus, consider a transition vector expressed as,
\begin{equation}\label{eq:transitionVector}
\bm{q} = (q_{CC}^1,q_{CD}^1,q_{DC}^1,q_{DD}^1;q_{CC}^2,q_{CD}^2,q_{DC}^2,q_{DD}^2).
\end{equation}
Each component, $q_{a\tilde{a}}^i$, of the transition vector gives the probability of the state of the stochastic game being $s_1$---the profitable state (recall that $b_1>b_2$)---in the next round, given the current state is $s_{i}$ and the current action profile is $(a,\tilde{a})$. We call a transition vector  deterministic if all $q_{a\tilde{a}}^i\in\{0,1\}$. The transition vector is symmetric if $q_{a\tilde{a}}^i = q_{\tilde{a}a}^i$. In this study, we find it sufficient and succinct to only consider the symmetric and deterministic transition vectors. Fig.~\ref{fig:cartoon} illustrates the setup built up to now.

There can still be $2^6$ such transition vectors. However, given that we have adopted the narrative of the ToC, the number of relevant vectors can be reduced to three as can be argued as follows. Our focussed interest in on the scenarios where defection by either or both the two interacting players results in a transition from the most beneficial state $s_{1}$ to the less beneficial state $s_{2}$. This means $q^1_{CD}=q^1_{DC}=q^1_{DD}=0$. Also, in the presence of mutual cooperation, $s_1$ should be sustained as such, i.e., $q^1_{CC}=1$, and  $s_{2}$ should transition to $s_{1}$, i.e., $q^2_{CC}=1$.  Thus, we arrive at a simplified expression for $\bm{q}$, viz., $(1,0,0,0;1,q^2_{CD},q^2_{CD},q^2_{DD})$, where we have put $q^2_{CD}=q^2_{DC}$, due to the aforementioned symmetry consideration. Obviously, only four possibilities remain. Next, we fix $q^2_{CD}\geq q^2_{DD}$ by demanding that if a single act of cooperation among the two players fails to facilitate a transition from $s_{2}$ to $s_{1}$, then such a transition remains unattainable even with mutual defection. Consequently, we are finally left with only three possible symmetric and deterministic transition vectors in line with the narrative of the ToC: $\bm{q_{00}}=(1,0,0,0;1,0,0,0)$, $\bm{q_{10}}=(1,0,0,0;1,1,1,0)$, and $\bm{q_{11}}=(1,0,0,0;1,1,1,1)$. \textcolor{black}{In the literature~\cite{Kleshnina2023}, games involving $\bm{q_{10}}$ and $\bm{q_{11}}$ are referred to as the timeout-with-conditional-return game and the timeout game, respectively.} 

{\color{black}Before proceeding further, it is useful to build a clear intuitive physical picture for the three transition vectors. To this end, we recall that  the framework we have adopted is a non-trivial simplification of multi-state resource (see Fig.~\ref{fig:cartoon}a). In the narrative of ToC, in addition of change in the self-renewing resource's state due to its growth rate ($r_i$), the act of cooperation is supposed to replenish the resource and that of defection is supposed to deplete the resource. The vector $\bm{q_{00}}$ can be thought to be representing a situation in which the growth rate of the resource vanishes or is very small; therefore, only mutual cooperation can effectively replenish the depleted resource. In this sense, the feedback is strongest when $\bm{q}=\bm{q_{00}}$, because maintaining the beneficial resource state requires cooperation from both players. The cases of $\bm{q_{10}}$ and $\bm{q_{11}}$ are more nuanced. These transition vectors correspond to situations where one act of cooperation (note $\bm{q_{10}}$)---and even no cooperation (note $\bm{q_{11}}$)--- in the depleted state, can be sufficient to return the resource to the beneficial state. At first glance a fact appears as an apparent inconsistency which can be starkly illustrated using the case of transition vector $\bm{q_{11}}$: One notes that while presence of even a single act of defection ($q_{CD}^1=q_{DC}^1=q_{DD}^1=0$) pushes the resource into deplete state $s_2$, defection by even both the players ($q_{DD}^2=1$) replenishes the resource to state $s_1$! This perplexing counter-intuitive asymmetric effect of defection is, however, easily resolved when one associates this $\bm{q_{11}}$ with an appropriate self-renewing resource: When the resource is almost replete ($m_i\sim K$), the logistic growth rate ($r_i$) is almost zero, so the negative effect of defection is maximally pronounced. However, when the resource is intermediately depleted ($m_{\rm thr}< m_i\ll K$), the growth rate is large enough (recall, $r_i<r_{i+1}$ $\forall i$) such that the resource may recover back to the replete state even in the absence of any cooperative action. Similar picture can be presented for $\bm{q_{10}}$ through analogous arguments.}

Finally, we elevate the setup to encompass an unstructured well-mixed population of fixed size $N$, assumed large enough. If a population is monomorphic such that all players are $\bm{p}$-strategied, then in the course of the stochastic game being played between every pair of  players, the average rate of cooperation per round, $\gamma(n_1,n_2,\bm{p},\bm{q})$, of a player is the measure of propensity for cooperation of the population. One must note that because of the presence of noisy channel and environment-action feedback, $\gamma$ is crucially dependent on its arguments. Mechanism for emergence and subsequent sustenance of cooperation is our main focus, and hence, the initial population is supposed to composed of individuals who always defect. The emergence of cooperation, then, definitely owes to mutation which is considered to be rare:  If a mutant appears in the population, then it either invades the resident population or becomes extinct, and till that happens, the probability of two or more consecutive mutations is negligible. Hence, the population consists of no more than two strategies at any time.  

As far the sustenance of cooperation is concerned, selection is of importance: We suppose that players do pairwise comparison~\cite{Traulsen2007} of expected payoffs to learn the opponent's strategy, or equivalently, one could just view it as a Moran process~\cite{Moran1958} driven by fitness difference. Once a player receives payoffs by interacting (through the stochastic game) with all other individuals in the population, its expected payoff , $\pi$, is compared with a randomly chosen role model (either resident or mutant) expected payoff, $\tilde{\pi}$; and the focal player switches to the role model's strategy with a probability $\left(1+{\rm exp}\left[-\beta(\tilde{\pi}-\pi) \right]\right)^{-1}$, where the non-negative parameter $\beta$ is the strength of the selection. At the end of this pairwise comparison process, the population is once again monomorphic: The mutants either become extinct or the new resident; in the latter case, $\gamma$ can change to a new value. Subsequently, another random mutant enters the population, and the aforementioned mutation-selection process repeats ad infinitum. Consequently, the propensity of cooperation in the population also evolves. 
As the resulting Markov chain describing the mutation-selection process is ergodic (see Appendix.~\ref{aAvgPayCoop}), $\hat{\gamma}$---the long-run time averaged cooperation rate---is independent of sample paths. 
For robust maintenance of cooperation, one requires $\hat{\gamma}$ to be near unity. 

Obviously, the set of states of this Markov chain is the set of strategies available to a player. At steady state, some strategy (denoted as $\bm{p}_{\rm m}$) will be the most probable (i.e., most recurringly appearing as resident) with frequency, say, $w_{\bm{p}_{\rm m}}$. We shall see later such strategies are quite important in understanding the system. Also, we point out that we consider the realistic phenomenon that a player would err~\cite{Boyd1989,Brandt2006}, say, with small probability $\epsilon$, while executing her intended action. This inclusion renders $\gamma$ of a monomorphic population independent of the players' initial moves~\cite{SigmundBook}.

Let us not forget that our game dynamics is intricately intertwined with the information channel $(n_{1},n_{2})$ which every population may not be utilizing with equal efficiency. To get a measure of the working of the information channel, given that the resident strategy is $\bm{p}$, it is useful to devise a measure called efficacy, $E$, defined as logarithm of the ratio of the mutual information to the capacity of the information channel. Mutual information of an (memoryless discrete) information channel~\cite{Shannon1948} is a measure of the rate of information transfer through the channel and is given by
\begin{eqnarray}
\begin{aligned}
I(n_1,n_2,\bm{p},\bm{q}) &\equiv I({\bm \alpha}_{\rm out}(n_1,n_2,\bm{p},\bm{q})|{\bm \alpha}_{\rm in}(n_1,n_2,\bm{p},\bm{q}))\\
& = H(\alpha_{\rm out})- \alpha_{\rm in}H(n_1)-(1- \alpha_{\rm in})H(n_2).
\end{aligned}
\end{eqnarray}
Here, $H(z)\equiv -z\log_2 z-(1-z)\log_2 (1-z)$ represents the entropy function; ${\bm \alpha}_{\rm in}\equiv\left(\alpha_{\rm in},(1-\alpha_{\rm in})\right)$ and ${\bm \alpha}_{\rm out}\equiv\left(\alpha_{\rm out},(1-\alpha_{\rm out})\right)$ are the probability distributions, respectively, over the input and the output sets of states ($s_1$ and $s_2$) of channel $(n_1,n_2)$. ${\alpha}_{\rm in}$ can be calculated by tracking with what frequency state $s_1$ (or $s_2$) appears in a sequence of the stochastic game. Subsequently, ${\alpha}_{\rm out}$ can be calculated using the error in perception due to the noisy channel: 
\begin{eqnarray}
\alpha_{\rm out} &=& (1-n_1) \alpha_{\rm in}+n_2 (1- \alpha_{\rm in}).
\end{eqnarray}
By definition, the channel capacity~\cite{Yang2022}
\begin{equation}
\begin{split}
K{(n_1,n_2)} &\equiv\sup_{{\bm \alpha}_{\rm in}}I({\bm \alpha}_{\rm out}(n_1,n_2,\bm{p},\bm{q})|{\bm \alpha}_{\rm in}(n_1,n_2,\bm{p},\bm{q}))\\
& =\log_2 \left(1+2^{\left[H(n_1)-H(n_2)\right]/\left[1-n_1-n_2\right]}\right)\\
&-\frac{1-n_2}{1-n_1-n_2}H(n_1) +\frac{n_1}{1-n_1-n_2}H(n_2).
\end{split}
\end{equation}
Capacity $K(n_{1},n_{2})$ of the channel is the maximum of the mutual information. Thus, by definition, the efficacy, 
\begin{equation}
E(n_1,n_2,\bm{p},\bm{q})={\rm log}_{10}\left(\frac{I(n_1,n_2,\bm{p},\bm{q})}{K(n_1,n_2)}\right),
\end{equation}
is always non-positive. Larger values of efficacy signify the greater ability of the resident strategy $\bm{p}$ in receiving information through the sensory channel attached to them; when $E\to0$, the resident population may be said to use the information channel almost up to its capacity.

The main goal for this paper to also see the effect of noise on $\gamma$. To this end, one may define
\begin{equation}\label{eq:BoNC}
\Delta\gamma(n_{1},n_{2},\bm{p},\bm{q})\equiv\gamma(n_{1},n_{2},\bm{p},\bm{q})-\gamma(0,0,\bm{p},\bm{q}).
\end{equation}
This measures the benefit of noise in sustaining cooperation. It is positive when the noise in channel helps enhancing the population's cooperation level. On somewhat similar note, we define a measure for benefit of noise in preventing the ToC:
\begin{equation}\label{eq:BoNa}
\Delta\alpha(n_{1},n_{2},\bm{p},\bm{q})\equiv \alpha_{\rm in}(n_{1},n_{2},\bm{p},\bm{q})-\alpha_{\rm in}(0,0,\bm{p},\bm{q}).
\end{equation}
This measure tells us about the propensity of avoiding the ToC. When it is positive, the state $s_1$ in the presence of noise is relatively more frequent when compared with the noiseless case. Like the cooperation rate, large time averages of $E$, $\Delta \gamma$, and $\Delta \alpha$ also evolve over time to their respective asymptotic values, viz., $\hat{E}$, $\Delta\hat{\gamma}$, and $\Delta\hat{\alpha}$.

\section{Results}
\begin{figure}[t!]
	\centering
	\includegraphics[scale=0.42]{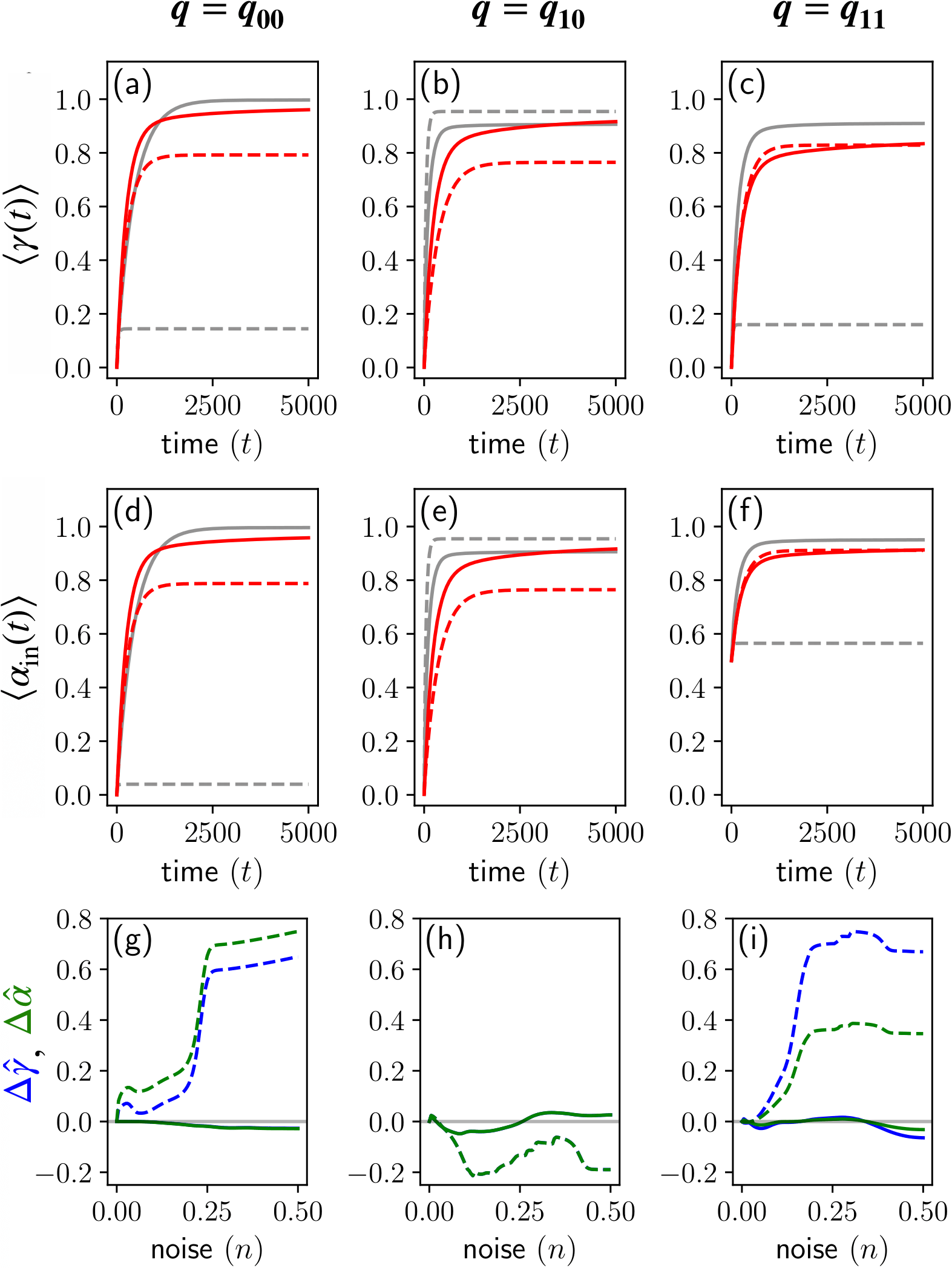}
	\caption{{\it Symmetric noisy channel mediated evolution of cooperation:} The first row (a-c) depicts the evolution of rate of cooperation, $\gamma(t)$, averaged over an ensemble of paths starting from unconditionally defecting resident population. The second row (d-f) exhibits corresponding evolutions of the frequency, $\alpha_{\rm in}$, of beneficial state. The first, the second, and the third columns, respectively, correspond to the outcomes of transition vectors $\bm{q_{00}}$, $\bm{q_{10}}$, and $\bm{q_{11}}$. In these plots, solid and dashed curves represent outcomes for populations with memory-$1$ and memory-$\frac{1}{2}$, respectively. The red and the grey colored curves in the first two rows, respectively, represent outcomes for cases where the information channel is maximally noisy $(n=0.5)$ and minimally noisy $(n=0)$. We observe that for the transition vectors $\bm{q_{00}}$ and $\bm{q_{11}}$, a noisy channel is beneficial for a population of memory-$\frac{1}{2}$ strategies, contrary to the case of the transition vector $\bm{q_{10}}$. This observation is quantified in the third row (g-i) that depicts the long-run time-averaged enhancement of cooperation, $\Delta\hat{\gamma}$ (blue curves), and of the probability of being in the most beneficial state, $\Delta\hat{\alpha}$ (green curves), for all possible symmetric noisy channels. For illustration purpose, we have fixed $N=100$, $b_{1} = 2.0$, $b_{2}=1.2$, $c=1.0$, $\epsilon=10^{-3}$, and $\beta=10$.}
	\label{fig:symmetric}
\end{figure}
\begin{figure*}[t!]
	\centering
	\includegraphics[scale=0.4]{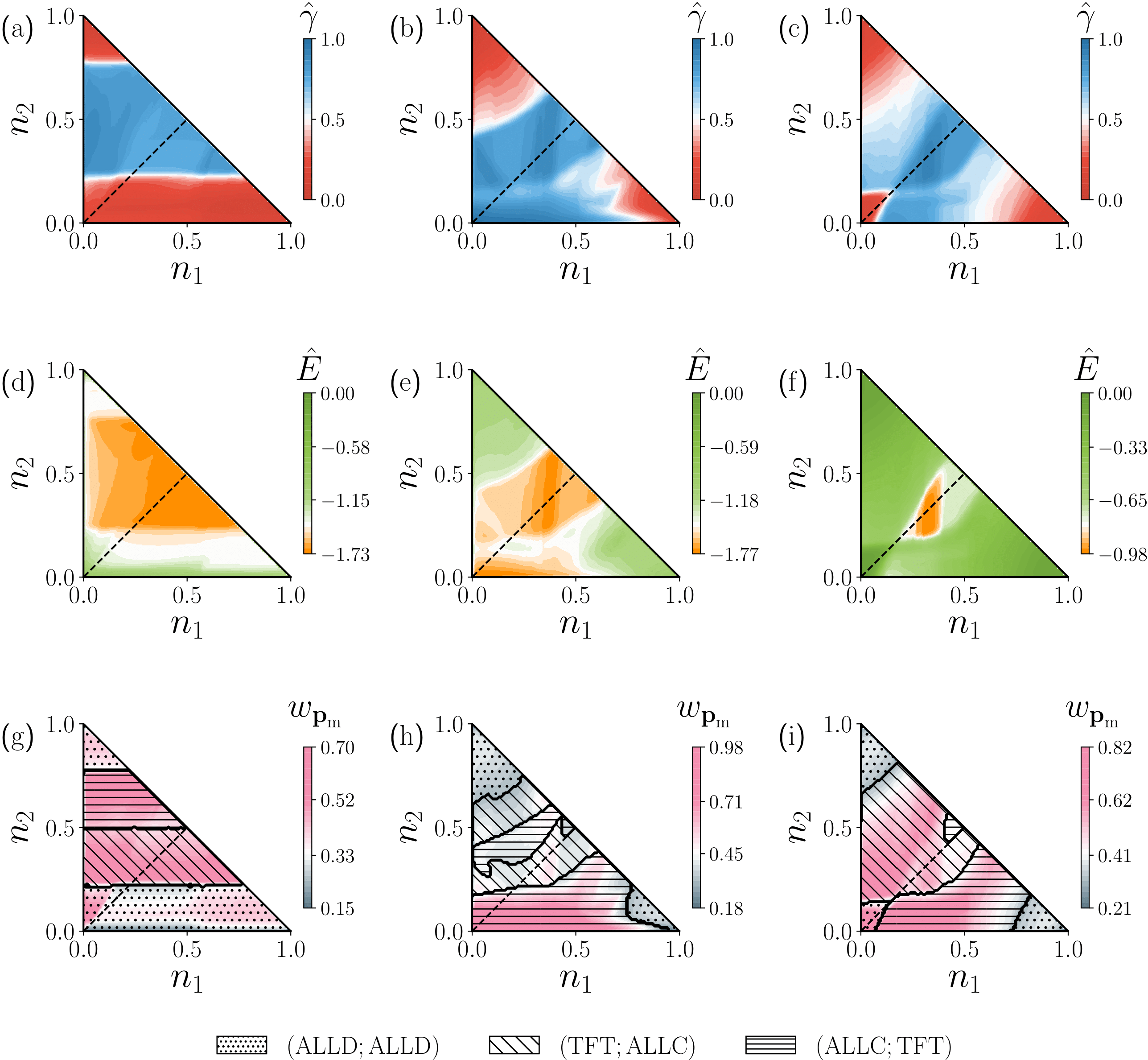}
	\caption{{\it Asymmetric noisy channel mediated evolution of cooperation and efficacy:} The first, second, and third columns represent the outcomes for transition vectors \(\bm{q_{00}}\), \(\bm{q_{10}}\), and \(\bm{q_{11}}\), respectively, for all possible binary asymmetric channels. In the first row (a-c), plots depict the long-run time-averaged cooperation rate, \(\hat{\gamma}\); in the second row (d-f), plots depict the long-run time-averaged efficacy, \(\hat{E}\); and in the third row (g-i), plots depict the maximally recurrent strategies \(\bm{p}_{\rm m}\) and their long run frequencies of recurrence \(w_{\bm{p}_{\rm m}}\) in the population. The dashed line \(n_{1}=n_{2}\) represents subset corresponding to the symmetric channels. For illustration purpose, we have fixed $N=100$, $b_{1} = 2.0$, $b_{2}=1.2$, $c=1.0$, $\epsilon=10^{-3}$, and $\beta=10$.} 
	\label{fig:channel}
\end{figure*}

Henceforth, unless otherwise specified, we consider only the pure strategies---either $2^8$ pure memory-1 strategies or $2^4$ reactive ones---transformed by noisy channel through Eq.~(\ref{eq:channelTrans}). To begin with we consider symmetric channel, i.e., $n_1=n_2=n$. As exhibited through the solid grey curves in Fig.~\ref{fig:symmetric}a-c, in the absence of any noise, the memory-1 strategy leads to asymptotically high level of cooperation even when the initial population is composed of individual who always defect irrespective of the environmental state. Given the association of the ToC via the choice of transition vectors---viz., $\bm{ q_{00}}$, $\bm{q_{10}}$, and $\bm{q_{11}}$ (respectively, corresponding to the three columns of Fig.~\ref{fig:symmetric}---it is assuring that there is accompanying prevalence of the beneficial state asymptotically; this can be concluded from the solid grey curves in Fig.~\ref{fig:symmetric}d-f showcasing the beneficial state's frequency. Both the reported $\gamma$ and  $\alpha_{\rm in}$ correspond to their average values. Interestingly, the conclusions are heavily dependent on the memory: In the case strategies are reactive and noise is absent, while the cooperation levels are comparatively abysmal for $\bm{ q_{00}}$ and $\bm{q_{11}}$, the cooperation level is better for $\bm{q_{10}}$. The fate of the beneficial environmental state follows the suit. A visual comparison of grey solid curves with respective dashed curves in  Fig.~\ref{fig:symmetric} validates this memory-dependence crystal clearly. 

The effect of noisy channel on the games is quite intriguing. Very first interesting aspect is that the asymptotic cooperation level achieved by a population in the presence of noisy-channel-induced highly incorrect perception (say, $n\to 1$) of the environmental states is same as the level achieved in the almost absence of the channel noise (say, $n\to 0$). More generally, one observes that the cooperation level and other related quantities are same whether the probability of incorrect perception of the state is $n$ or $1-n$. This is apparent, in rather general form, from Eq.~(\ref{eq:channelTrans}) for any binary channel (not necessarily symmetric): All one has to do is replace  $n_j$ by $n'_j\equiv1-n_{\bar{j}}$. In the light of the fact that we are considering only pure strategies, i.e., $p^j_{a\tilde{a}}$'s in the right hand side of Eq.~(\ref{eq:channelTrans}) can only be $0$ or $1$, following conclusion is self-evident: The set of $2^8$ strategies in the presence of channel $(n_1,n_2)$ is mapped bijectively to itself under the transformation $(n_1,n_2)\to(n'_1,n'_2)$; moreover, the strategy---(0,0,0,0;0,0,0,0), always defect, irrespective of the environmental state---maps to itself. In other words, since the starting point of the evolutionary dynamics and the set of mutants remains unchanged under the transformation, the dynamical outcome at any point $(n_1,n_2)$ in the $n_1$-$n_2$ space must be equivalent to that at the point $(n'_1,n'_2)$, which is a mirror reflection of point $(n_1,n_2)$ about the line $n_1+n_2=1$. The same conclusion straightforwardly carries over to the subset of strategies that constitute reactive strategies.

The effects of noise on $\bm{q_{00}}$-game, $\bm{q_{10}}$-game and $\bm{q_{11}}$-game has a common feature: When the memory-1 strategies are in action, the noise does not change the stationary cooperation level and the stationary frequency of beneficial state much. This is clear from the third row of Fig.~\ref{fig:symmetric} where $\Delta\hat{\gamma}$ (blue solid curve) and $\Delta\hat{\alpha}$ (green solid curve) hover around zero for all values of $n$. However, for the reactive strategies, the effect of channel noise is significant. For high enough channel noise, $\Delta\hat{\gamma}$ (blue dashed curve) and $\Delta\hat{\alpha}$ (green dashed curve) are quite high and positive for $\bm{q_{00}}$-game and $\bm{q_{11}}$-game; in $\bm{q_{10}}$-game, however, $\Delta\hat{\gamma}$ and $\Delta\hat{\alpha}$ are not as high (but still they are higher than the memory-1 case) and are negative. In other words, channel noise is deleterious in $\bm{q_{10}}$-game. These conclusions are illustrated using the specific case of $n=0.5$ in the first and the second rows of Fig.~\ref{fig:symmetric} where saturation of the cooperation level and frequency of the beneficial state are, respectively, depicted.

Evidently, the class of reactive strategies are more interesting in the context of binary noisy channel. Hence, let us expand our domain of investigation to asymmetric information channels, $(n_1,n_2)$ with $n_1\ne n_2$, in this class. As per the discussion above, it suffices to limit the study to the cases for which $n_{1}+n_{2}\leq1$, without any loss of generality. 

A look at the equilibrium cooperation rate $\hat{\gamma}$ and the efficacy $\hat{E}$ (see first and the second rows of Fig.~\ref{fig:channel}, respectively) for all the three games reveals one very important feature: Whenever non-zero noise renders the evolutionary system significantly cooperative (blue regions in Fig.~\ref{fig:channel}a--c), the long-term efficacy in processing the information about the state by the individuals in the population plunges (yellow regions in Fig.~\ref{fig:channel}d--f).  In other words, the strategies that facilitates receiving of information about the states with a transmission rate far less than the capacity of the information channel are chosen by evolution more frequently. After all, the mutual information, by definition, represents the uncertainty about channel input~\cite{djordjevic2022quantum} resolved on observing channel output---it depicts the amount of information per symbol transmitted by the channel. Here, the set of symbols is $\{s_1,s_2\}$. 

While the inverse connection between cooperation and efficacy is intriguing, some more interesting features can be extracted from Fig.~\ref{fig:channel}. Firstly, in $\bm{q_{00}}$-game and $\bm{q_{11}}$-game, the system is unable to become cooperative without the presence of the noise; whereas, in $\bm{q_{10}}$-game, the system achieves a significant cooperation level even without noise. Secondly, in all the three games, low $n_2$ (less error in perceiving deplete state) but high $n_1$ (high error in perceiving beneficial state) leads to defection (see the bottom right red corners in the first row of Fig.~\ref{fig:channel}) . Of course, same happens for high $n_2$ but low $n_1$  leads to defection (see the top left red corners in the first row of Fig.~\ref{fig:channel}). Thirdly, in the presence of intermediate noise levels, the cooperation is brought forth by the channel. 

In this context, it is of insight to consider a scenario where there exists a probability $e$ that players become unable to determine the state of the stochastic game. In such instances of indeterminacy regarding the game state, it is natural that the players, guided by the `principle of insufficient reason'~\cite{keynes1921treatise}, would randomly assume a state of the stochastic game and employ strategy accordingly. The effect of such a binary asymmetric channel with erasure probability $e$ is analogous to that of a binary asymmetric channel $(n^{e}_1,n^{e}_2)$ without any erasure, where $n^e_{i}\equiv n_{i}(1-e)+\tfrac{e}{2}$. Consequently, we note that with increasing erasure of information, cooperation may be established because as the probability $e$ approaches unity, $n^e_{i}$ tends towards $\tfrac{1}{2}$ regardless of the values of $n_{i}$. At $(\tfrac{1}{2},\tfrac{1}{2})$, the channel has about maximum cooperation as seen Fig.~\ref{fig:channel}.

\section{Mechanism}
Comprehending the afore-discussed emergent behaviours requires us to understand the relative dominance of various strategies; after all, whether the population would be cooperative or not, is decided by the cooperation level of the strategies that appear as resident-strategies comparatively more recurrently as the population undergoes mutation-selection process to reach a steady state. 

In the case of reactive strategies, one fortunately has a manageable number of them which can be compactly expressed by $4$-tuple representation $\bm{p}=(p_{C}^1,p_{D}^1;p_{C}^2,p_{D}^2)$ where the subscripts represent the action of only the opponent in the last round. Since, in our study, each component $p^i_{\tilde{a}}$ is allowed to take values $0$ or $1$, each $\bm{p}$ can be represented as a $4$-digit binary number of a decimal integer from 0 to 15. For example, the strategy of unconditional cooperation (ALLC) in any state can be represented as $({\rm ALLC};{\rm ALLC})=\bm{p}=(1,1;1,1)$. Just like ALLC (always cooperate), using the conventional~\cite{Baek2016,SigmundBook} meaning of TFT (tit-for-tat) and ALLD (always defect), three other strategies of  immediate interest are conveniently represented as $({\rm ALLC};{\rm TFT})=(1,1;1,0))$, $({\rm TFT};{\rm ALLC})=(1,0;1,1)$, and $({\rm ALLD};{\rm ALLD})=(0,0;0,0)$.

Now, for all three transition vectors, we plot (Fig.~\ref{fig:channel}g--i) the maximally recurrent strategies ($\bm{p}_{\rm m}$) and their equilibrium fractions, ($w_{\bm{p}_{\rm m}}$) of recurrences for all the channels in $n_{1}$-$n_{2}$ space. We observe that, throughout $n_{1}$-$n_{2}$ space for all the three games, these are: $({\rm ALLD};{\rm ALLD})$, $({\rm TFT};{\rm ALLC})$, and $({\rm ALLC};{\rm TFT})$, exclusively. A generic observation is that for all the three games, the population becomes cooperative when either $({\rm TFT};{\rm ALLC})$ or $({\rm ALLC};{\rm TFT})$ becomes the most recurrent one. Otherwise, when  $({\rm ALLD};{\rm ALLD})$ becomes the most recurrent, the cooperation becomes unstable as expected. 

\begin{figure}[t]
	\centering
	\includegraphics[scale=0.35]{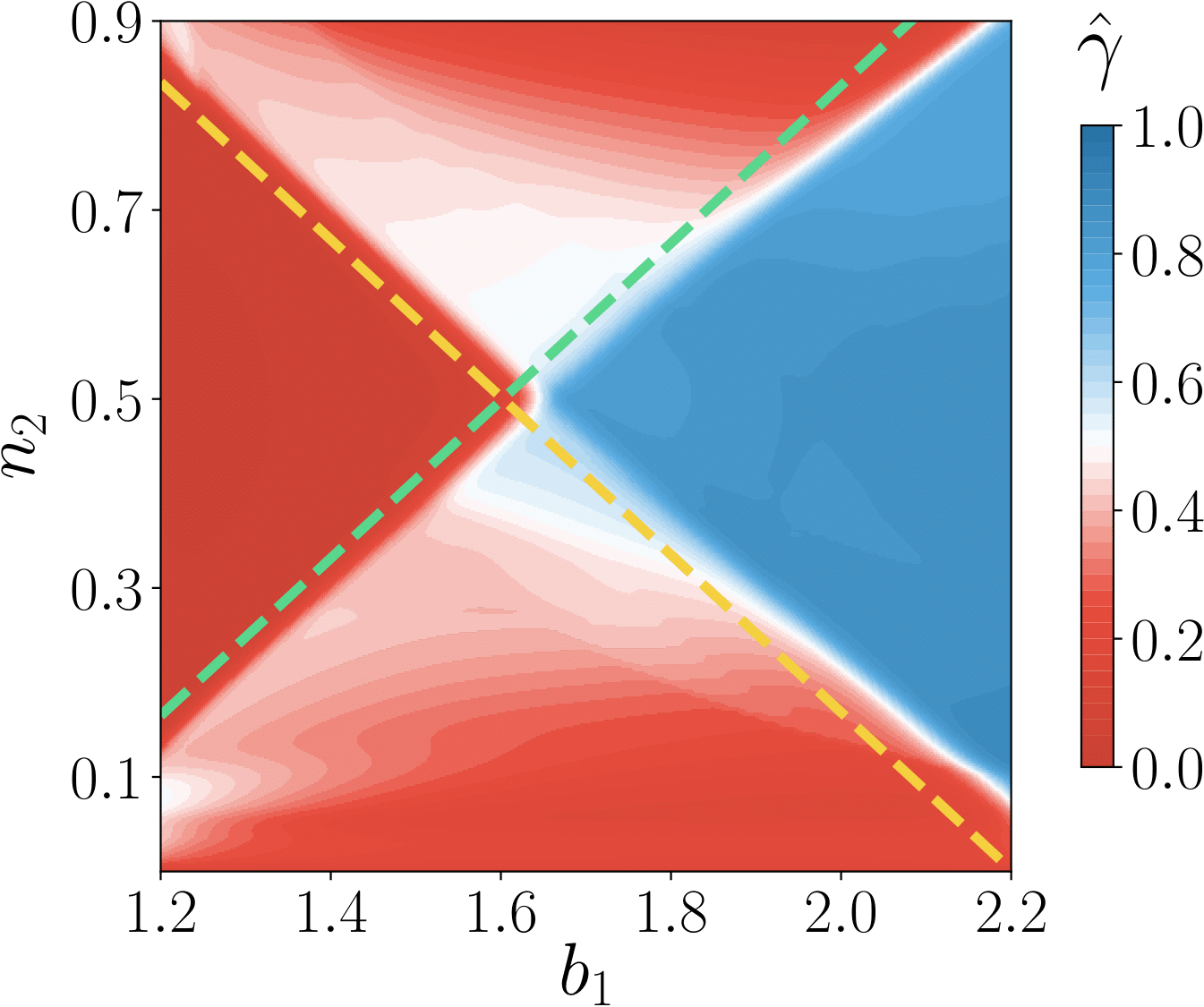}
	\caption{{\it Only three reactive strategies decide sustenance of the cooperation:} In the plot of $b_{1}$ vs. $n_{2}$, the region above the analytically found yellow dashed line is where payoff of $({\rm TFT};{\rm ALLC})$ against $({\rm TFT};{\rm ALLC})$ is greater than the payoff of $({\rm ALLD};{\rm ALLD})$ against $({\rm TFT};{\rm ALLC})$; and the region below the analytically found green dashed line is where payoff of $({\rm ALLC};{\rm TFT})$ against $({\rm ALLC};{\rm TFT})$ is greater than the payoff of $({\rm ALLD};{\rm ALLD})$ against $({\rm ALLC};{\rm TFT})$. Note that the numerically found common (blue) region of cooperation is in line with the analytical prediction.  Here, $N=100$, $b_{2}=1.2$, $c=1.0$, $n_{1}=0.1$, $\epsilon=10^{-3}$, and $\beta=10$.}
	\label{fig:stability}
\end{figure}

In order to hone our intuition about how distribution of $w_{\bm{p}_{\rm m}}$ helps to explain the sustenance of cooperation, let us consider the simplest case: $n_1=n_2=0.5$ in any of the three stochastic games. In this case, the strategies $({\rm TFT};{\rm ALLC})$ and $({\rm ALLC};{\rm TFT})$ become effectively identical in the light of transformation~(\ref{eq:channelTrans}). This is clear in Fig.~\ref{fig:channel}g--i where at $(n_1,n_2)=(0.5,0.5)$, the  horizontal lined region ($\bm{p}_{\rm m}$=({\rm TFT};{\rm ALLC})) and the slanted lined region ($\bm{p}_{\rm m}$=({\rm ALLC};{\rm TFT})) touch. There the population equilibrates to maintain a highly cooperative state as is clear from the corresponding point in Fig.~\ref{fig:channel}a--c. Both these strategies either choose to cooperate regardless of the previous actions of the opponent or mimic the opponent's last action. As both are self-cooperative, they swiftly restore the state of the stochastic game, facilitating mutual cooperation in the most advantageous state. Dangerous to the population is $({\rm ALLD};{\rm ALLD})$ strategy---note how it persists in dotted regions in Fig.~\ref{fig:channel}g--i leading to absence of cooperation, thence to the ToC. Why it does not spread elsewhere, including the case $n_1=n_2=0.5$ under discussion? Well, the two self-cooperative strategied players counter any defection by opponent with defection with a probability of $0.5$ in any given round, regardless of the game's state. This feature renders these strategies less prone to exploitation by the defectors compared to unconditional cooperation. 

We know that~\cite{Hilbe2018} in the stochastic game (in the absence of noisy channel), the evolution of cooperation owes to the fact that defectors face dual disadvantage: They cannot improve the state of stochastic game to benefit from there and additionally, they receive occasional defection from the reciprocal opponents. Noisy channel's presence accentuates the disadvantage further. As we have described above, the noise---owing to transformation~(\ref{eq:channelTrans})---changes the pure strategies to mixed strategies, unless a pure strategy, e.g., $({\rm ALLD};{\rm ALLD})$, is composed of identical strategies in both the states. This mixing of strategies gives rise to reciprocating behaviour which resists exploitation by the defectors, incurring further loss to the defectors and hence facilitating evolution of cooperation in the population.

Of course, in scenarios with arbitrary $n_1$ and $n_2$ in any of the three stochastic games, more detailed picture of strategies' distribution is needed. However, the importance of the three maximally recurrent strategies may be highlighted by establishing that how good an approximation it is to merely consider only these three strategies among all sixteen pure strategies. To this end, as an illustration, consider the particular case of $\bm{q_{00}}$-game as seen in Fig.~\ref{fig:channel}a: It is observed that the population exhibits cooperative behaviour regardless of the value of $n_{1}$, provided the condition $\tfrac{1}{5}\lessapprox n_{2}\lessapprox\tfrac{4}{5}$ is satisfied. This is very simply explained by demanding that the payoff fetched by $({\rm ALLD};{\rm ALLD})$ against both $({\rm TFT};{\rm ALLC})$ and $({\rm ALLC};{\rm TFT})$ is less than what these self-cooperating strategies fetch against themselves. Please see Fig.~\ref{fig:stability} and Appendix.~\ref{aAnalyticalCond} in this context.

The slight mismatch between this analytical prediction of the region of cooperation as seen in blue colour in Fig.~\ref{fig:stability} is naturally attributed to the presence of other thirteen strategies. Although the mismatch is minor, the influence of other strategies, especially $({\rm ALLC};{\rm ALLC})$ and $({\rm TFT};{\rm TFT})=(1,0;1,0)$, in establishing cooperation starting from $({\rm ALLD};{\rm ALLD})$ is paramount. While $({\rm ALLC};{\rm ALLC})$ acts as stepping stone for $({\rm ALLD};{\rm ALLD})$ to invade $({\rm TFT};{\rm ALLC})$ and $({\rm ALLC};{\rm TFT})$, $({\rm TFT};{\rm TFT})$ is the catalyst for beating $({\rm ALLD};{\rm ALLD})$ to result in a cooperative state. The interplay between these five strategies is effectively the minimal reduced Markov chain that can closely reproduce Fig.~\ref{fig:channel} (see Appendix.~\ref{aMinimalRep}) which is not reproduced at all by not considering $({\rm ALLC};{\rm ALLC})$ and $({\rm TFT};{\rm TFT})$.

{\color{black} The effectiveness of stochastic games in generating cooperation was originally done~\cite{Hilbe2018} using state-independent memory-1 strategies. In that setting, the strategy Win-Stay--Lose-Shift (WSLS) plays a crucial role as it has an intrinsic error-correcting mechanism unlike the reactive strategy, TFT. Recall that a WSLS strategy in Prisoner's Dilemma~\cite{Hilbe2018} (in the absence of any stochastic switching), by construction, means that the player would stick to her action in next round if the opponent cooperated in the last round; otherwise, she would shift to the alternative action. Thus, when any of the two WSLS players playing a sequence of $C$'s, does an execution error $D$, the subsequent round has both the players playing $D$ and therefore, the round after that shift back to the original sequence of both playing $C$'s. In contrast, when two TFT-players are playing with each other, an error $D$ in the ongoing sequence of all $C$'s leads to a situation where both the TFT players are trapped in a never-ending sequence of alternating $C$ and $D$---thus, the single execution error degrades the overall cooperative state. Thus, it is natural to wonder: If the error-correcting feature of WSLS is responsible for sustaining cooperation in stochastic games with state-independent memory-1 strategies, then is there some analogous error-correcting feature present in our setup of stochastic games with reactive strategies? The answer is in affirmative: Actually, the desired feature is a natural result of the combination of noisy channel and state-dependence of strategies, as we explain below.

Let us take the cases of afore-discussed two of maximally recurrent state-dependent reactive strategies identified in our analysis, both of which comprise two reactive components---TFT in one state and ALLC in the other. In the presence of noise, the recognition of the game state frequently becomes faulty, leading players to respond as if they were in the wrong dilemma. When such a misperception occurs, a player momentarily switches from the TFT strategy (which enforces reciprocity) to the ALLC strategy (which embodies forgiveness). Thus, in the most noisy symmetric case, i.e., $n = \frac{1}{2}$, both players either copy the action chosen by the opponent last round (i.e., owing to TFT) or cooperate (owing to ALLC), each with equal probability. Naturally, alternating cooperation-defection cycles or sequences of mutual defection---resulting in reducing cooperation level---triggered by a small execution error are not feasible because of the effective `error-correction' due to erroneously perceiving ALLC through the noisy channel.

However, the reader should note that the entire mechanism of error correction---whether in WSLS or in the two state-dependent strategies---operates effectively only when the execution error is small. So far, we have considered a small probability of execution error $\epsilon = 10^{-3}$ in the paper. It is, however, important to assess the robustness of cooperation against the magnitude of execution error. It is obvious that when $\epsilon$ is as high as $0.5$, all strategies lose their distinct identities and effectively become random: Players cooperate or defect with equal probability, regardless of context. This randomness has two inevitable consequences---the long-run frequencies of recurrence of all strategies in the population become equal and the population consequently loses its cooperative state. Thus, we can reasonably predict that as the magnitude of the execution error increases, the overall level of cooperation declines in the population. To verify this, we numerically analyze $\epsilon$ over a wide range $\left[10^{-5},10^{-1}\right]$ to find our intuition validated (see Fig.~\ref{fig:execError} in Appendix~\ref{app:ep}).}

\section{Discussion}
In biological and social contexts, cooperation, which is crucial for the smooth functioning of systems, often faces challenges~\cite{Hamilton1964,Axelrod1981,Bendor1997,Lehmann2006} from the forces of natural selection, particularly in competitive surroundings. While cooperation benefits groups by fostering the creation and sharing of social resources, it remains vulnerable to exploitation leading to the ToC, where individuals reap the rewards without contributing to communal efforts~\cite{hardin1968commons,ostrom1999coping, Rankin2007}. However, the evolutionary puzzle of cooperation finds resolution through specific mechanisms~\cite{Nowak2006}, facilitating the emergence and enduring stability of cooperative behaviour. Here we have investigated the intricate interplay between noisy information processing and the evolution of cooperation within the paradigm of stochastic games. While the effect of evolving environments on evolutionary dynamics has been very recently explored~\cite{Hilbe2018} using the theory of stochastic games, the exploration of noisy channel's effect on such stochastic games using information-theoretic concepts is the novelty of this paper. The idea of noisy sensory channels adds a realistic layer of complexity to the evolutionary dynamics of populations engaging in social dilemmas.  

Our setup is simplest non-trivial one: Directly reciprocating players repeatedly interact while perceiving binary environmental state through a binary asymmetric channel and the state of the environment changes based on the players action. This setup renders mathematical tractability of the question in hand. Moreover, we could keep our focus more on the simpler setting of reactive strategies because of the herein discovered fact that noisy perception can enhance or deteriorate cooperation significantly in populations employing reactive strategies but its impact on memory-$1$ strategy populations is subdued. Subsequently, through combination of analytical and numerical analyses, the setup demonstrates that noisy perception of the environment can lead to higher levels of cooperation by improving the population's ability to adapt to changing environmental conditions. This phenomenon is observed across three all different transition schemes---which fit the narrative of the ToC---of the stochastic games. 

Furthermore, the efficacy measure offers a quantitative assessment of the population's utilization of noisy information channels, indicating the extent to which individuals exploit imperfect information for cooperative ends. The investigation into symmetric and asymmetric information channels reveals the channels' nuanced effects on the dynamics governing the evolution of cooperation. While the extent of cooperative outcomes unsurprisingly depends on the type of stochastic game, it is quite intriguing that the scenarios of high cooperative state coincide with low efficacies. The qualitative reason behind this may be comprehended as follows: One realizes that in the two self-cooperative strategies, $({\rm TFT};{\rm ALLC})$ and $({\rm ALLC};{\rm TFT})$, it is the ${\rm TFT}$ part that fends off $({\rm ALLD};{\rm ALLD})$ from invading. Low efficacy implies that the information about the true states is not correctly perceived, thereby leading the individuals to believe themselves to be in both the states. Due to this, the individuals with either of the self-cooperative strategies, are not stuck with only ${\rm ALLC}$; rather, being misled by noise, they play ${\rm TFT}$ often. Consequently, the self-cooperative strategies prevail and cooperation is established.

The framework expounded in this paper on noisy channel mediated mitigation of the ToC is conducive to many other interesting inquiries, e.g., one could leave the domain of repetition of simultaneous one-shot games to address how effective noise is in alternating repeated games~\cite{Nowak1994}.  While details are presented in the accompanying Appendix.~\ref{aInteractions}, we mention here that the effectiveness of noise in sustaining cooperation while mitigating the ToC is a remarkable common feature even in the alternating stochastic games. \textcolor{black}{The effect of demographic stochasticity~\cite{Wang2023} in stochastic game played in a population of non-constant size is another worthy research avenue.} Moreover, there may be other kinds of noisy channels in action, e.g., there may be an additional noisy channel that accounts for the fact that players may occasionally misread the action chosen by the opponent. Our study provokes investigation of such other channels. We conclude by posing one more question, which we would like to take up in future: How does the noisy perception of environment affect the evolution of cooperation and mitigation of the ToC when the players are engaged in indirect reciprocity~\cite{Nowak1998}?

\subsection*{Data, Materials, and Software Availability}
All the computer codes used in this paper were written in C++, Python 3.0, and Mathematica 14.0, and they are available \href{https://github.com/samratsohel/Noisy-information-channels-mediated-prevention-of-the-tragedy-of-the-commons.git}{online}.

\subsection*{Acknowledgments}
Authors are thankful to Mayank Pathak and Supratim Sengupta for helpful discussions. SC acknowledges the support from SERB (DST, govt. of India) through project no. MTR/2021/000119.

\appendix

\section{Appendix}
We present here supporting information essential for gaining deeper insight into the model discussed in the main text. In Section~\ref{aErrorExecution}, we detail how incorporation of the probability of execution error $\epsilon$ by the player within our model has been done. Section~\ref{aInteractions} delves into the mathematical framework of Markov chain, which consists of states representing various combinations of the stochastic game states and action profiles. Specifically, Subsection~\ref{aInteractionsInf} addresses the scenario where interactions between players are repeated infinitely many times, while Subsection~\ref{aInteractionsDisc} explores cases where the probability of subsequent rounds occurring is less than unity (notably, this scenario is not discussed in detail in the main text). Section~\ref{aExpectedPayCoop} elaborates on the computation of the expected payoff and cooperation in a two-player repeated stochastic game. The mathematical formulation of imitation dynamics in large and finite populations is covered in Section~\ref{aEvolutionaryDynamics}. The method for computing the long-run time-averaged cooperation rate is discussed in Section~\ref{aAvgPayCoop}. Section~\ref{aEvolutionCoop} focuses on ensemble-averaged cooperation rates at any given time. The calculation of the capacity of an information channel is briefly addressed in Section~\ref{aMeasures}. We argue that the population size used in this study is sufficiently large in Section~\ref{aconvergenceTest}. In Section~\ref{aAnalyticalCond}, we derive a condition for the stability of cooperation, specifically for the transition vector $\bm{q_{00}}$ in the case of simultaneous moves without the influence of future interactions. The key strategies driving the outcomes discussed in the main text and argument for these strategies to be important are highlighted in Section~\ref{aMinimalRep}. The results corresponding to alternating moves by the players are examined in Section~\ref{aResAlt} and compared with that of the simultaneous games. Finally, we conclude with Section~\ref{aShadowScheme} which provides a comparison of the outcomes of simultaneous and alternating moves in the presence of the shadow of the future.

\subsection{Error in execution}\label{aErrorExecution}
Players may make execution errors~\cite{Boyd1989,Brandt2006} with probability $\epsilon$ when implementing their intended actions. Mathematically, the effective strategy of a player with initial strategy $\bm{p}=(p_{a\tilde{a}}^j)$ is transformed by the relation~\cite{Baek2016,Hilbe2018,Kleshnina2023}:
\begin{equation}\label{aeq:err}
p_{a\tilde{a}}^j\to\epsilon+(1-2\epsilon)p_{a\tilde{a}}^j.
\end{equation}
In repeated games where errors are present, the long-run dynamics are not dependent on the players' initial moves~\cite{SigmundBook}. For this study, we assume the execution error probability $\epsilon$ to be small, set at $\epsilon=10^{-3}$. Additionally, in a noisy communication channel, a player's strategy $\bm{p}=(p_{a\tilde{a}}^j)$, in the absence of execution errors, undergoes a transformation as follows:
\begin{equation}\label{aeq:chan}
p_{a\tilde{a}}^j\to (1-n_{j})p_{a\tilde{a}}^j+n_{j}p_{a\tilde{a}}^{\bar{j}}.
\end{equation}
Transformations~\ref{aeq:err} and~\ref{aeq:chan} are commutative. Thus, with a noisy channel $(n_{1}, n_{2})$ and an execution error rate $\epsilon$, an element of the effective strategy transforms as follows:
\begin{equation}\label{aeq:channelTrans}
p_{a\tilde{a}}^j\to\epsilon+(1-2\epsilon)\big[p_{a\tilde{a}}^j+n_{j}(p_{a\tilde{a}}^{\bar{j}}-p_{a\tilde{a}}^j)\big].
\end{equation}
As it should be, in the limit as $n_{j}\to0$, transformation~\ref{aeq:channelTrans} converges to~\ref{aeq:err}, representing the case of error-free perception; whereas, in the limit as $\epsilon\to0$, transformation~\ref{aeq:channelTrans} converges to~\ref{aeq:chan}, representing the case of error-free execution.

\subsection{Two-player repeated interaction}\label{aInteractions}
\begin{figure*}[t!]
	\centering
	\includegraphics[scale=0.25]{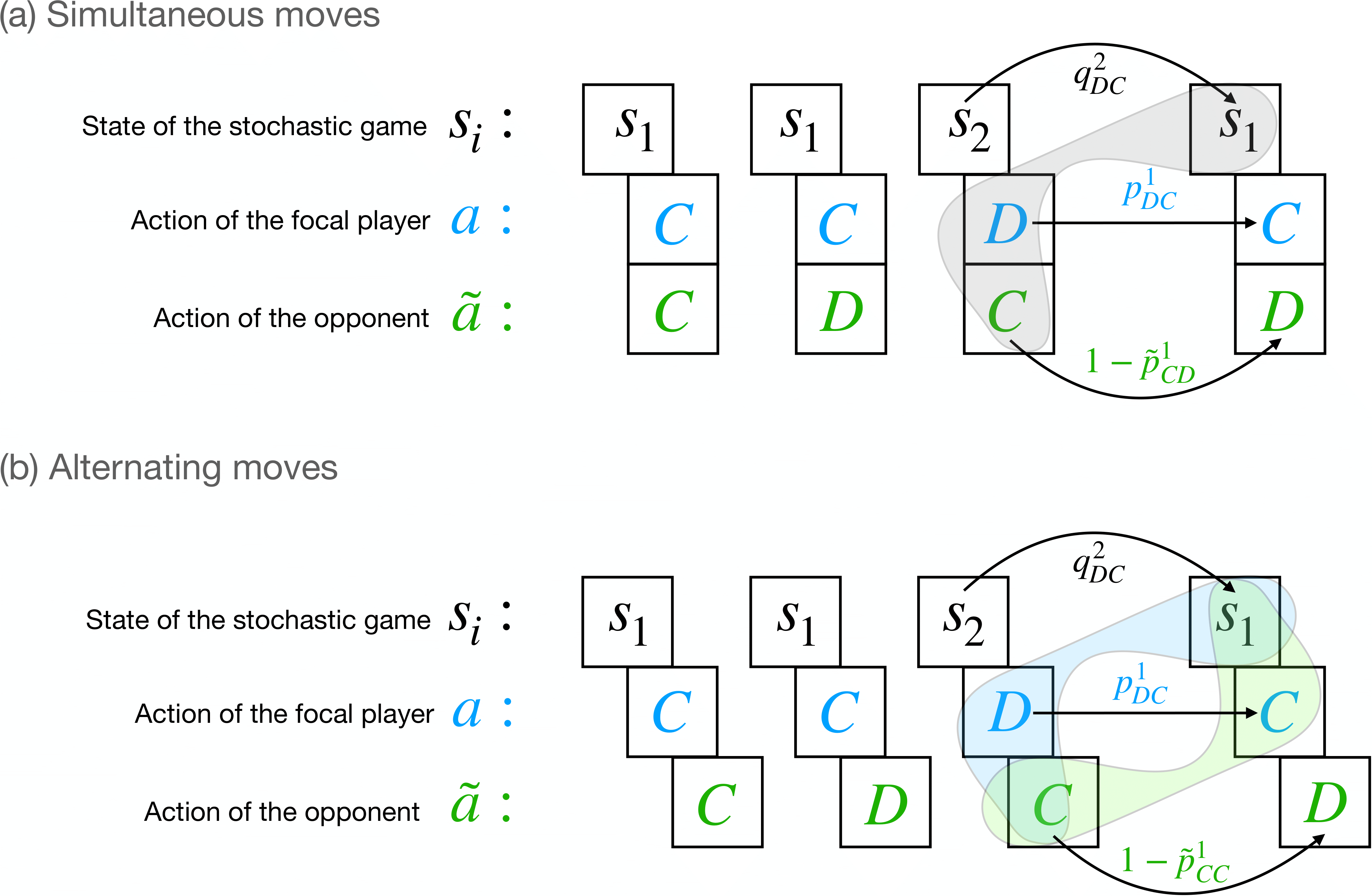}
	\caption{{\it Representative sequences of the state in the stochastic game, action $a$ of the focal player with strategy $\bm{p}$, and action $\tilde{a}$ of the opponent with strategy $\tilde{\bm{p}}$, with simultaneous (a) and alternating (b) move schemes:} For both schemes, the Markov chain starts from state $\omega_0 = (s_1, C, C)$. After two rounds, the game reaches $\omega_2 = (s_2, D, C)$. The probability that the state of the stochastic game is $s_1$, regardless of the move scheme, is given by $P(\omega_3 = (s_i = s_1, a, \tilde{a}) \mid \omega_2) = q^2_{DC}$. Upon transitioning to state $s_1$, the focal player observes her last action as $D$ and the opponent's as $C$ (gray contour), leading them to cooperate with probability $p^1_{DC}$. Similarly, the opponent, observing her last action as $C$ and the focal player's as $D$ (gray contour), cooperates with probability $\tilde{p}^1_{CD}$. In the alternating move scheme, the focal player, acting as the leader, observes her last action as $D$ and the opponent's as $C$ (blue contour), and thus cooperates with probability $p^1_{DC}$. The opponent, seeing her last action as $C$ and the focal player's as $C$ (green contour), since the opponent acts after the leader, cooperates with probability $\tilde{p}^1_{CC}$.}
	\label{afig:schemes}
\end{figure*}

The smallest unit of the model, a pair of players, say, player-$1$ and player-$2$, interact repeatedly. In a round, both the players either choose their actions simultaneously or in alternating fashion (see Fig.~\ref{afig:schemes}). A Markov chain can describe this interaction. Given that the effective strategies (after transformation~\ref{aeq:channelTrans}) of player-$1$ and player-$2$ are ${\boldsymbol{p}}$ and $\tilde{{\boldsymbol{p}}}$, respectively, and the state of the environment has been modelled to take two values---$s_1$ and $s_2$, this Markov chain consists of eight possible states: Any state can be represented by $\omega = (s_i,a,\tilde{a})$.  The transition probability of moving from state $\omega = (s_i,a,\tilde{a})$ to state $\omega' = (s_{i'},a',\tilde{a}')$ is a product of three factors as depicted below:
\begin{equation}
m_{\omega,\omega'} = x\cdot y\cdot \tilde{y}.
\end{equation}
The first factor is due to the transition between states. So this factor is calculated using the elements of transition vector ${\boldsymbol{q}}$ as follows,
\begin{equation}
x =
\begin{cases}
q^i_{a\tilde{a}}  & \text{if $s_{i'}=s_1$},\\
1-q^i_{a\tilde{a}} & \text{if $s_{i'}=s_2$}.
\end{cases}
\end{equation}
The other two factors, $y$ and $\tilde{y}$, correspond to the conditional probabilities that the focal player and her opponent respectively choose the actions prescribed in $\omega'$ is state $s_{i'}$. So, these factors are calculated using the elements of strategies ${\boldsymbol{p}}$ and $\tilde{\boldsymbol{p}}$ as follows:
\begin{equation}
y =
\begin{cases}
p^{i'}_{a\tilde{a}}  & \text{if $a'=C$},\\
1-p^{i'}_{a\tilde{a}} & \text{if $a'=D$}.
\end{cases}
\end{equation}
Unlike ${y}$, $\tilde{y}$ has dissimilar relations in the cases of simultaneous moves and alternating moves. For simultaneous moves by the players 
\begin{equation}
\tilde{y} =
\begin{cases}
\tilde{p}^{i'}_{\tilde{a}a}  & \text{if $\tilde{a}'=C$},\\
1-\tilde{p}^{i'}_{\tilde{a}a} & \text{if $\tilde{a}'=D$};
\end{cases}
\end{equation}
whereas, for alternating moves by the players (considering focal player to be the leader and the opponent to be the follower),
\begin{equation}
\tilde{y} =
\begin{cases}
\tilde{p}^{i'}_{\tilde{a}a'} & \text{if $\tilde{a}'=C$},\\
1-\tilde{p}^{i'}_{\tilde{a}a'} & \text{if $\tilde{a}'=D$}.
\end{cases}
\end{equation}
By calculating all possible $m_{\omega,\omega'}$, we obtain transition matrix ${\sf M}({\boldsymbol{p}},\tilde{\boldsymbol{p}})\equiv (m_{\omega,\omega'})$, of the Markov chain. 

\subsubsection{Infinitely repeated games}\label{aInteractionsInf} If the game is repeated infinitely many times, transition vector ${\boldsymbol{q}} \neq (1,~1,~1,~1,~0,~0,~0,~0)$ and execution error $\epsilon\notin\{0,1\}$, then the corresponding Markov chain is ergodic and irrespective of the initial state, the probabilities over the states always converge to unique limiting values. In that case, ${\sf M}({\boldsymbol{p}},\tilde{\boldsymbol{p}})$ has a unique left eigenvector ${\boldsymbol{v}}=(v_{a\tilde{a}}^i)$ with unit eigenvalue. Each entry $v_{a\tilde{a}}^i$ represents the probability of the state $\omega = (s_i,a,\tilde{a})$ of the Markov chain in the long run. In the main text, we have considered exactly same scenario where the focal player and the opponent interact for an infinite number of rounds without any discounting. In this case, we emphasize the fact that, the probability $\delta$ of the next round occurring is assumed to be unity.

\subsubsection{Shadow of the future}\label{aInteractionsDisc}
Here we additionally explore the situations where the probability of the upcoming round occurring is not unity, i.e., $0<\delta<1$. Hence, with shadow of future, the $k$th round occurs with probability $\delta^{k-1}$. Initially if the state vector is $\bm{u}$, we will get a sequence of states as $\bm{u},\bm{u}{\sf M},\bm{u}{\sf M}^2,\bm{u}{\sf M}^3,\cdots$. So, instead of directly computing the left eigenvector of the transition matrix ${\sf M}$, we derive the wighted averaged state vector as follows:
\begin{eqnarray}\label{aeq:shadowSim}
\bm{v} &=& \frac{\sum_{k=1}^{\infty}\bm{u}(\delta{\sf M})^{k-1}}{\sum_{k=1}^{\infty}\delta^{k-1}}= (1-\delta)\bm{u}({\sf I}-\delta{\sf M})^{-1}
\end{eqnarray}
Here, ${\sf I}$ is an identity matrix of size $8\times8$. 

In the case of $\delta<1$, the initial specification matters in predicting the final one. For concreteness, let us assume that a player with pure memory-$\tfrac{1}{2}$ strategy $\bm{p}=(p_1,p_2;p_3,p_4)$ cooperates in the very first round of the stochastic game, which starts from beneficial state $s_{1}$, with a probability $p_1$, assuming the opponent to be a cooperative one. Given that the strategies of the focal player and the opponent are $\bm{p}$ and $\tilde{\bm{p}}$, respectively, the elements of the vector $\bm{u}=(u^{i}_{a\tilde{a}})$ can be calculated as the product of two factors, $x$ and $\tilde{x}$ as detailed below:

First for the simultaneous game case, we find
\begin{equation}\label{aeq:ffvo}
x =
\begin{cases}
{p}_{1} & \text{if ${a}=C$ and $i=1$},\\
1-{p}_{1} & \text{if ${a}=D$ and $i=1$},\\
0 & \text{if $i=2$};
\end{cases}
\end{equation}
and
\begin{equation}\label{aeq:ffvo2}
\tilde{x} =
\begin{cases}
\tilde{p}_{1} & \text{if $\tilde{a}=C$ and $i=1$},\\
1-\tilde{p}_{1} & \text{if $\tilde{a}=D$ and $i=1$},\\
0 & \text{if $i=2$}.
\end{cases}
\end{equation}

Next for the alternating game case, we realize that calculation is a bit subtle: The initial state vector has an asymmetry caused by the fact that the focal player may either be leader or follower. We call the initial state vector as $\bm{u^{\rm leader}}$. In this case the factor $x$ remains same as written in Eq.~\ref{aeq:ffvo}. However, the factor $\tilde{x}$ changes to
\begin{equation}
\tilde{x} =
\begin{cases}
\tilde{p}_{1} & \text{if $\tilde{a}=C$, $a=C$, and $i=1$},\\
1-\tilde{p}_{1} & \text{if $\tilde{a}=D$, $a=C$, and $i=1$},\\
\tilde{p}_{2} & \text{if $\tilde{a}=C$, $a=D$, and $i=1$},\\
1-\tilde{p}_{2} & \text{if $\tilde{a}=D$, $a=D$, and $i=1$},\\
0 & \text{if $i=2$}.
\end{cases}
\end{equation}
This is different from Eq.~\ref{aeq:ffvo2} because the opponent no longer needs to assume that the focal player has cooperated---it can go by what the focal player has played (either $C$ or $D$) immediately before her.

In case the focal player is the follower, using similar arguments, for the initial state vector, now denoted as $\bm{u^{\rm follower}}$, we get
\begin{equation}
\tilde{x} =
\begin{cases}
\tilde{p}_{1} & \text{if $\tilde{a}=C$ and $i=1$},\\
1-\tilde{p}_{1} & \text{if $\tilde{a}=D$ and $i=1$},\\
0 & \text{if $i=2$};
\end{cases}
\end{equation}
and
\begin{equation}
x =
\begin{cases}
p_{1} & \text{if $a=C$, $\tilde{a}=C$, and $i=1$},\\
1-p_{1} & \text{if $a=D$, $\tilde{a}=C$, and $i=1$},\\
p_{2} & \text{if $a=C$, $\tilde{a}=D$, and $i=1$},\\
1-p_{2} & \text{if $a=D$, $\tilde{a}=D$, and $i=1$},\\
0 & \text{if $i=2$}.
\end{cases}
\end{equation}
Futhermore, assuming that the focal player can be leader or follower with equal probability, we calculate the average initial state vector as $\bm{u}=\left[\bm{u^{\rm leader}}+\bm{u^{\rm follower}}\right]/2$. After calculating $\bm{u}$, we can calculate $\bm{v}$ using Eq.~\ref{aeq:shadowSim}.

Henceforth, whatever follows is true for all values of $\delta$ with the caveat that for $\delta=1$, $\bm{v}$ should be interpreted as the limiting distribution of  ${\sf{}M}({\boldsymbol{p}},\tilde{\boldsymbol{p}})$, whereas for $0<\delta<1$, Eq.~\ref{aeq:shadowSim} dictates the meaning of $\bm{v}$.

\subsection{Expected payoff and cooperation rate of a repeated game}\label{aExpectedPayCoop}
As the entries $v^i_{a\tilde{a}}$ of vector ${\boldsymbol{v}}$ gives the probabilities with which the players observe the state $\omega = (s_i,a,\tilde{a})$, we can compute the average payoff of the players using the knowledge of the payoff corresponding to each state of the Markov chain. So, for a given transition matrix ${\sf M}({\boldsymbol{p}},\tilde{\boldsymbol{p}})$, the focal player's expected payoff can be found to be
\begin{equation}
\pi({\boldsymbol{p}},\tilde{\boldsymbol{p}}) = \sum_{i\in\{1,2\}}\left[b_i\left(\sum_{a\in\{C,D\}}v^i_{aC}\right)-c\left(\sum_{\tilde{a}\in\{C,D\}}v^i_{C\tilde{a}}\right)\right].
\end{equation}
We compute the average cooperation rate by the pair of the players as
\begin{equation}
\gamma({\boldsymbol{p}},\tilde{\boldsymbol{p}}) = \sum_{i\in\{1,2\}}\left[v^i_{CC}+\frac{(v^i_{CD}+v^i_{DC})}{2}\right].
\end{equation}
Note, for a infinitely repeated game $\pi({\boldsymbol{p}},\tilde{\boldsymbol{p}})\neq\pi(\tilde{\boldsymbol{p}},{\boldsymbol{p}})$, in general. However, equality $\gamma({\boldsymbol{p}},\tilde{\boldsymbol{p}})=\gamma(\tilde{\boldsymbol{p}},{\boldsymbol{p}})$ is always true. The quantity $\gamma({\boldsymbol{p}},\boldsymbol{p})$ is called the self-cooperation rate of the strategy $\bm{p}$.

\subsection{Evolutionary dynamics}\label{aEvolutionaryDynamics}
Our model assumes an unstructured population of fixed size $N$. Players engage in pairwise comparisons~\cite{Traulsen2007} of average payoffs to adopt new strategies under the condition of rare mutations. Each player receives payoffs by interacting with all other individuals in the population. A random focal player compares her payoff $\pi$ with a randomly chosen role model's payoff $\tilde{\pi}$ and switches to the role model's strategy with probability $\left(1 + \exp\left[-\beta(\tilde{\pi} - \pi)\right]\right)^{-1}$. The parameter $\beta$, representing selection strength, is non-negative, with higher values increasing selection strength. The process continues as the mutant strategy either fixates or becomes extinct, followed by the introduction of a new random mutant, repeating this cycle indefinitely. Let us now formalize this process mathematically. Initially, all players adopt the same resident strategy ${\boldsymbol{p}}_{\rm res}$, until a random mutation prompts one player to transition to an alternative strategy ${\boldsymbol{p}}_{\rm mut}$. The fate of this mutant strategy, whether it goes extinct or becomes fixed, hinges on its average payoff compared to a player employing the resident strategy as well as chance. At any intermediate juncture, if the population contains $k$ mutants employing strategy ${\boldsymbol{p}}_{\rm mut}$ and $(N-k)$ individuals adhering to the resident strategy ${\boldsymbol{p}}_{\rm res}$, the expected payoffs for a resident and a mutant can be, respectively, computed to be
\begin{widetext}
\begin{eqnarray}
\pi_{\rm res}(k) &=& \frac{N-k-1}{N-1}\cdot\pi({\boldsymbol{p}}_{\rm res},{\boldsymbol{p}}_{\rm res})+\frac{k}{N-1}\cdot\pi({\boldsymbol{p}}_{\rm res},{\boldsymbol{p}}_{\rm mut}),\\
\pi_{\rm mut}(k) &=& \frac{N-k}{N-1}\cdot\pi({\boldsymbol{p}}_{\rm mut},{\boldsymbol{p}}_{\rm res})+\frac{k-1}{N-1}\cdot\pi({\boldsymbol{p}}_{\rm mut},{\boldsymbol{p}}_{\rm mut}).
\end{eqnarray}
\end{widetext}
With these average payoffs, we use to standard results available in literature to write the fixation probability of the mutant strategy within the resident population as:
\begin{widetext}
\begin{equation}\label{aeq:fixation}
\rho({\boldsymbol{p}}_{\rm res},{\boldsymbol{p}}_{\rm mut}) = \frac{1}{1+\sum_{i=1}^{N-1}\prod_{k=1}^i\exp\left(-\beta\left[\pi_{\rm mut}(k)-\pi_{\rm res}(k)\right]\right)}.
\end{equation}
\end{widetext}
Subsequently, either with probability $\rho({\boldsymbol{p}}_{\rm res},{\boldsymbol{p}}_{\rm mut})$ the mutant strategy infiltrates the resident population, supplanting the current strategy, or with probability $1-\rho({\boldsymbol{p}}_{\rm res},{\boldsymbol{p}}_{\rm mut})$, the mutant goes extinct and the resident strategy remains unchanged. This evolutionary process repeats indefinitely, with new mutants appearing randomly in the population.

\subsection{Long run time-averaged cooperation and payoff}\label{aAvgPayCoop}
By iterating this afore-discussed evolutionary process over $\tau$ time steps, commencing with a resident strategy ${\boldsymbol{p}}_0$, we generate a sequence of $\tau+1$ resident strategies, denoted as $\mathcal{X}(0,\tau)=({\boldsymbol{p}}_0,{\boldsymbol{p}}_1,{\boldsymbol{p}}_2,\cdots,{\boldsymbol{p}}_{\tau})$. Utilizing this sequence, we can compute the long-term average cooperation rate and payoff of the population, respectively, to be
\begin{eqnarray}
\hat{\gamma} &=& \lim_{\tau\to\infty}\frac{1}{\tau+1}\sum_{t=0}^{\tau}\gamma({\boldsymbol{p}}_t,{\boldsymbol{p}}_t),\\
\hat{\pi} &=& \lim_{\tau\to\infty}\frac{1}{\tau+1}\sum_{t=0}^{\tau}\pi({\boldsymbol{p}}_t,{\boldsymbol{p}}_t).
\end{eqnarray}
Due to the ergodic nature of the evolutionary process for any finite $\beta$, these time averages exist and are independent of the initial resident population strategy, ${\boldsymbol{p}}_0$. If the strategy set, from which mutant strategies are drawn randomly, has a finite cardinality, these average quantities can be computed exactly. The aforementioned sequence can be treated as a Markov chain, possessing a number of states equal to the cardinality of the strategy set. This finite Markov chain is irreducible and aperiodic, thus, admitting a unique limiting distribution. Each possible resident strategy ${\boldsymbol{p}}$ corresponds to a state of this Markov chain. Mutants drawn from the strategy set enter the population uniformly at random. At any given moment, the probability of transitioning from the current resident strategy ${\boldsymbol{p}}$ to the next resident strategy $\tilde{{\boldsymbol{p}}}$ is expressed as:
\begin{equation}
r({\boldsymbol{p}},\tilde{\boldsymbol{p}}) =
\begin{cases}
\frac{\rho({\boldsymbol{p}},\tilde{\boldsymbol{p}})}{|\mathcal{P}|} & \text{if ${\boldsymbol{p}}\neq\tilde{\boldsymbol{p}}$},\\
1-\sum_{\tilde{\boldsymbol{p}}(\neq\tilde{\boldsymbol{p}})}\frac{\rho({\boldsymbol{p}},\tilde{\boldsymbol{p}})}{|\mathcal{P}|} & \text{if ${\boldsymbol{p}}=\tilde{\boldsymbol{p}}$}.
\end{cases}
\end{equation}
Here, $\mathcal{P}$ denotes the strategy set from which mutants are drawn. By computing the unique left eigenvector ${\boldsymbol{w}}=(w_{\boldsymbol{p}})$ of the matrix ${\sf R} \equiv (r({\boldsymbol{p}},\tilde{\boldsymbol{p}}))$, we get the unique stationary distribution of the Markov chain. Leveraging this knowledge of the vector ${\boldsymbol{w}}$, we can precisely compute the long-run time-averaged cooperation rate and payoff as follows:
\begin{eqnarray}\label{aeq:hatgamma}
\hat{\gamma} &=& \sum_{{\boldsymbol{p}}\in\mathcal{P}}w_{\boldsymbol{p}}\gamma({\boldsymbol{p}},{\boldsymbol{p}}),\\
\hat{\pi} &=& \sum_{{\boldsymbol{p}}\in\mathcal{P}}w_{\boldsymbol{p}}\pi({\boldsymbol{p}},{\boldsymbol{p}}).\label{aeq:hatpi}
\end{eqnarray}
This method for computing the averages is computationally more efficient and accurate compared to obtaining a long-time average from a single simulation. For all cases discussed in this study, we utilize this approach to calculate the long-run time-averaged cooperation and payoff in the population. 

\subsection{Evolution of cooperation}\label{aEvolutionCoop}
To understand the evolutionary trajectory of cooperation level $\gamma(t)$ within a population starting the strategy of unconditional defection or `{{always defect}}' (ALLD), denoted as $\boldsymbol{p}_0$, we undertake the averaging of cooperation level starting from $t=0$ to $t=\tau$ process across $n$ distinct realizations of the trajectory $\mathcal{X}(0,\tau)$, as expressed below:
\begin{widetext}
\begin{equation}
\langle {\Gamma}(0,\tau)\rangle_{n}= (\gamma(\boldsymbol{p}_0,\boldsymbol{p}_0),\langle \gamma(\boldsymbol{p}_1,\boldsymbol{p}_1)\rangle_{n},\langle\gamma( \boldsymbol{p}_2,\boldsymbol{p}_2)\rangle_{n},\ldots,\langle\gamma(\boldsymbol{p}_{t},\boldsymbol{p}_{t})\rangle_{n}).
\end{equation}
\end{widetext}
Here, ${\Gamma}(0,\tau)$ is a path comprising of cooperation rates $\gamma(t)$, corresponding to a path $\mathcal{X}(0,\tau)$. By evaluating the limit as $n$ tends to infinity, $\lim_{n\to\infty}\langle\gamma(\boldsymbol{p}_{t},\boldsymbol{p}_{t})\rangle_{n}=\langle{\gamma}(t)\rangle$, under the condition that all trajectories originate from $\boldsymbol{p}_0$, we derive the vector ${\boldsymbol{w}}(t)={\boldsymbol{w}}(0)\cdot{\sf R}^{t}$, where ${\boldsymbol{w}}(0)$ represents a row vector with unity assigned to the element corresponding to the strategy $\boldsymbol{p}_0$ and zero to all other elements. Each element of the row vector ${\boldsymbol{w}}(t)$ provides the probability of a specific strategy being prevalent within the population at time $t$. This understanding of the vector ${\boldsymbol{w}(t)}$ enables us to exactly compute the average cooperation rate at any time step $t$ as depicted below:
\begin{eqnarray}\label{aeq:avgamma}
\langle{\gamma}(t)\rangle_{n\to \infty} &=& \sum_{\boldsymbol{p}\in\mathcal{P}} w_{\boldsymbol{p}}(t)\gamma(\boldsymbol{p},\boldsymbol{p}),
\end{eqnarray}
We note that ergodicity property of the Markov chain under consideration ensures that $\langle{\gamma}(t\to \infty)\rangle_{n\to \infty}=\hat{\gamma}$. So, irrespective of the initial state the long-run path averaged cooperation level is equal to the long-run time averaged cooperation level computed from a sample path.

\subsection{Capacity of the information channel}\label{aMeasures}
To gauge the effectiveness of information processing by a resident strategy $\bm{p}$ given the information channel is $(n_{1},n_{2})$, we introduced a metric called `efficacy' $E$ in the main text. It is the base-$10$ logarithm of the ratio of the mutual information $I(n_{1},n_{2},\bm{p},\bm{q})$ to the capacity $K(n_{1},n_{2})$ of the information channel. The capacity of a channel, denoted $K{(n_1,n_2)}$, represents the maximum mutual information over all possible input marginal probability distributions $\left[\alpha_{\rm in},1-\alpha_{\rm in}\right]$. 

In order to find the expression of $K{(n_1,n_2)}$, we calculate the derivative of $I$ with respect to $\alpha_{\rm in}$ and we get,
\begin{widetext}
\begin{equation}
\frac{\partial I}{\partial \alpha_{\rm in}}=(1-n_1-n_2)\log_2\left[\frac{1}{\alpha_{\rm in}(1-n_1-n_2)+n_2}-1\right]-\left[H(n_1)-H(n_2)\right].
\end{equation} 
\end{widetext}
By putting $\tfrac{\partial I}{\partial \alpha_{\rm in}}|_{\alpha_{\rm in}=\alpha_{\rm in}^*}=0$ and solving for $\alpha_{\rm in}^*$ we get,
\begin{equation}
\alpha_{\rm in}^* = \frac{1}{1-n_1-n_2}\left[\frac{1}{2^{\frac{H(n_1)-H(n_2)}{1-n_1-n_2}}+1}-n_2\right].
\end{equation}
Note that we obtain only one extremum at the value of $\alpha_{\rm in}^*$. Given that $I$ is a non-negative quantity with two zeros at $\alpha_{\rm in} \in \{0,1\}$, we can assert that the only extremum, $\alpha_{\rm in}^*$, is indeed a maximum. By substituting $\alpha_{\rm in}^*$ into $I$, we obtain the channel capacity, which can be expressed in a simplified form in terms of a channel $(n_1, n_2)$ as:
\begin{widetext}
\begin{equation}
\begin{split}
K{(n_1,n_2)} =& \log_2 \left(1+2^{\left[H(n_1)-H(n_2)\right]/\left[1-n_1-n_2\right]}\right)-\frac{1-n_2}{1-n_1-n_2}H(n_1) +\frac{n_1}{1-n_1-n_2}H(n_2).
\end{split}
\end{equation}
\end{widetext}
It is important to note that the capacity of a channel, unlike mutual information, is inherent to the channel $(n_1,n_2)$ and remains independent of the resident strategy and the transition vector under consideration. Physically, the channel capacity of an information channel denotes the maximum rate at which information can be reliably transmitted through the channel.

\begin{figure*}[t!]
	\centering
	\includegraphics[scale=0.5]{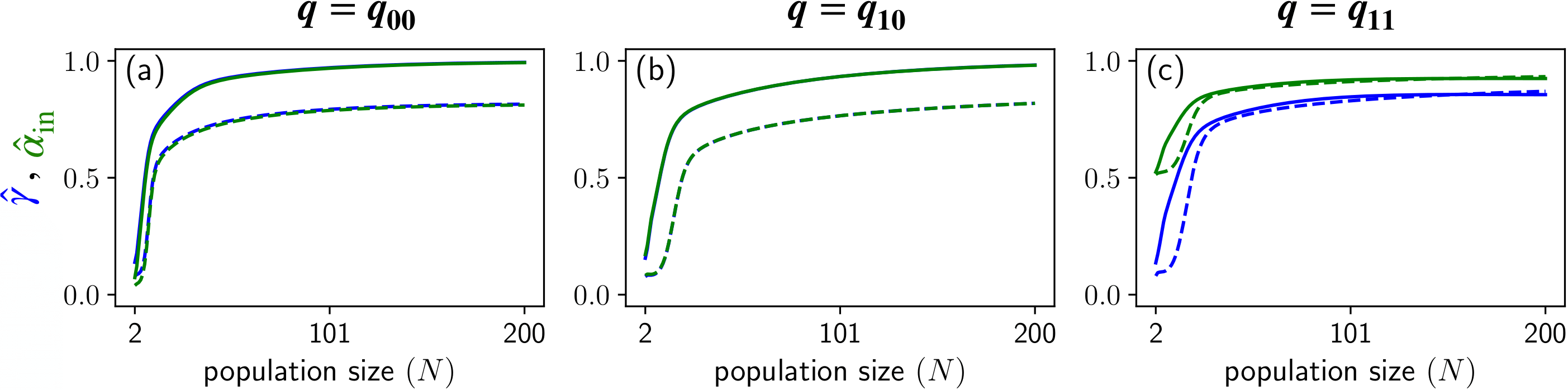}
	\caption{{\it Convergence test with different population size $N$ with symmetric noise $n$ and simultaneous moves by the player:} The blue curves represent the long-run time-averaged cooperation level $\hat{\gamma}$, while the green curves represent the long-run time-averaged probability of the beneficial state $\hat{\alpha}_{\rm in}$. The solid and dashed curves correspond to the outcomes for populations using memory-1 strategies and memory-$\frac{1}{2}$ strategies, respectively. Figures a, b, c represents outcomes for transition vectors $\bm{q_{00}}$, $\bm{q_{10}}$, and $\bm{q_{11}}$ respectively. We fixed the parameters, $b_{1} = 2.0$, $b_{2} = 1.2$, $c = 1.0$, $\epsilon = 10^{-3}$, $\beta = 10$, and $n = 0.5$.}
	\label{afig:ct}
\end{figure*}

\begin{figure*}[t!]
	\centering
	\includegraphics[scale=0.5]{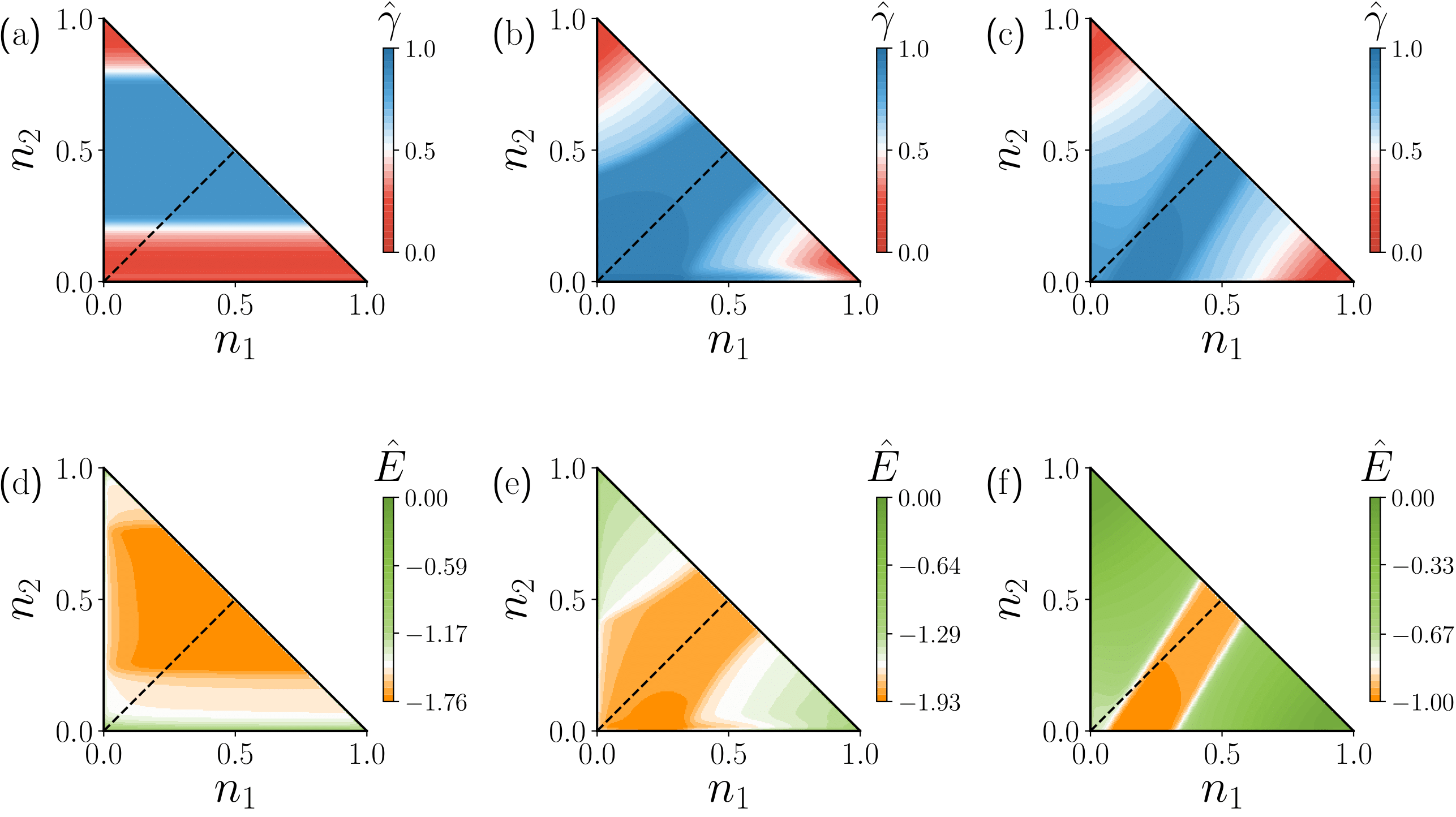}
	\caption{{\it Long-run time-averaged cooperation rate and efficacy of a memory-$\tfrac{1}{2}$ population with simultaneous moves by the players (considering the reduced Markov chain ${\sf R}_{\rm reduced}$ discussed in section~\ref{aMinimalRep}):} The first, second, and third columns represent the outcomes for transition vectors $\bm{q}=\bm{q_{00}}$, $\bm{q}=\bm{q_{11}}$, and $\bm{q}=\bm{q_{10}}$, respectively. The figures in the first (a, b, and c) and second (d, e, and f) rows show the color plots of equilibrium cooperation rate $\hat{\gamma}$ and efficacy $\hat{E}$ in the population, respectively. The points on the dashed line $n_{1}=n_{2}$ represent all the symmetric channels. The parameter values are fixed as follows: $N=100$, $b_{1} = 2.0$, $b_{2}=1.2$, $c=1.0$, $\epsilon=10^{-3}$, and $\beta=10$.}
	\label{afig:cr}
\end{figure*}

\subsection{Convergence test}\label{aconvergenceTest}
In this study, we have chosen a population size of $N = 100$ in our numerical simulations, a choice that warrants justification which we give now. We calculated the long-run time-averaged cooperation rate, $\hat{\gamma}$, and the probability of the beneficial state, $\hat{\alpha}_{\rm in}$, for population sizes ranging from $N = 2$ to $N = 200$ across all three stochastic game types: $\bm{q_{00}}$, $\bm{q_{10}}$, and $\bm{q_{11}}$, as discussed in the main text (see Fig.~\ref{afig:ct}). Our findings indicate that once the population size surpasses $N = 25$, the equilibrium outcomes increase only marginally with increasing $N$. In contrast, for very small population sizes, the equilibrium outcomes initially have relatively low values and show a rapid increase as the population size grows. Beyond this initial rapid increase, the values plateau, with only marginal increases observed, and the comparative results for memory-1 versus memory-$\tfrac{1}{2}$ strategies remain consistent. Additionally, the qualitative nature of the outcomes remains unchanged for population sizes beyond $N = 25$. Specifically, we chose a population size of $N = 100$ to ensure comparability with previous studies~\cite{Nowak2006,Kleshnina2023}, which also use the same population size. This consistency allows for a more accurate and direct comparison of results across different studies. 

\subsection{Analytical condition of stability of cooperation}\label{aAnalyticalCond}
For all three transition vectors we observe that for the maximally recurrent strategies there are three possibilities, i.e., $({\rm ALLD};{\rm ALLD})$, $({\rm TFT};{\rm ALLC})$, and $({\rm ALLC};{\rm TFT})$ (see the third row of Fig.~3). We contend that the changes observed in stability of cooperation can be comprehensively understood using the concept of `evolutionary stable strategy' (ESS)~\cite{SigmundBook}. An ESS is defined as a resident strategy that cannot be invaded by a small fraction of a mutant strategy, assuming the population is infinitely large. In our analysis, the population size is $N=100$, which is deemed sufficiently large. In a theoretically infinite population, if the resident strategy is an ESS, it categorically prevents invasion by mutants. Conversely, in large but finite populations, an ESS resident strategy exhibits increased resistance to being overtaken by a mutant. Thus, for the transition vector $\bm{q_{00}}$, we assert that when the noise strength is $n_{2} > \tfrac{1}{5}$, the resident strategy $({\rm TFT};{\rm ALLC})$ qualifies as an ESS, and similarly, when the noise strength is $n_{2} < \tfrac{4}{5}$, the resident strategy $({\rm ALLC};{\rm TFT})$ remains an ESS against the invasion by mutant $({\rm ALLD};{\rm ALLD})$, provided the population size is effectively infinite.

Acknowledging the fact that $\pi(({\rm ALLD};{\rm ALLD}),({\rm ALLD};{\rm ALLD}))$ is consistently greater than $\pi(({\rm TFT};{\rm ALLC}),({\rm ALLD};{\rm ALLD}))$ irrespective of the noise strengths, for $({\rm TFT};{\rm ALLC})$ to qualify as an ESS against the mutant $({\rm ALLD};{\rm ALLD})$, it is necessary that $\pi(({\rm TFT};{\rm ALLC}),({\rm TFT};{\rm ALLC})) > \pi(({\rm ALLD};{\rm ALLD}),({\rm TFT};{\rm ALLC}))$ (this condition satisfied by the region above the yellow line in Fig.~\ref{fig:stability} in the main text). Analytically, we can calculate these payoffs and express the conditions in terms of noise strength $n_{2}$, for the transition vector $\bm{q_{00}}$, as follows:
\begin{equation}\label{aeq:ineqAbove}
n_{2}>\frac{1}{2f_{2}(b_1,b_2,\epsilon)}\left[\sqrt{f_{1}(b_1,b_2,\epsilon)}+f_{3}(b_1,b_2,\epsilon)\right].
\end{equation}
Here, functions $f_{1}(b_1,b_2,\epsilon)=-3 \epsilon ^4 (b_1-b_2)^2+8 \epsilon ^3 (b_1-b_2)^2+2 \epsilon ^2 (b_1-b_2) (-3 b_1+4 b_2+1)-4 b_2 \epsilon  (b_1-b_2)+(b_1-1)^2$, $f_{2}(b_1,b_2,\epsilon)=(1-2 \epsilon ) ((1-\epsilon ) \epsilon  (b_1-b_2)+b_2)$, and $f_{3}(b_1,b_2,\epsilon)=-\left((1-\epsilon )^2 (b_1-b_2)\right)+b_2+1$. Similarly, as $\pi(({\rm ALLD};{\rm ALLD}),({\rm ALLD};{\rm ALLD}))$ is always greater than $\pi(({\rm ALLC};{\rm TFT}),({\rm ALLD};{\rm ALLD}))$ for being $({\rm ALLC};{\rm TFT})$ an ESS with respect to mutant $({\rm ALLD};{\rm ALLD})$ it is required that $\pi(({\rm ALLC};{\rm TFT}),({\rm ALLC};{\rm TFT}))>\pi(({\rm ALLD};{\rm ALLD}),({\rm ALLC};{\rm TFT}))$ (this condition satisfied by the region below the green line in Fig.~\ref{fig:stability} in the main text). Analytically, we can compute these payoffs and reduce the condition in terms of noise strength $n_{2}$, for the transition vector $\bm{q_{00}}$, as follows,
\begin{equation}\label{aeq:ineqBelow}
n_{2}<\frac{1}{2f_{2}(b_1,b_2,\epsilon)}\left[\sqrt{f_{1}(b_1,b_2,\epsilon)}+f_{4}(b_1,b_2,\epsilon)\right].
\end{equation}
Here, function $f_{4}(b_1,b_2,\epsilon)=-\left((5-4 \epsilon ) \epsilon ^2 (b_1-b_2)\right)+(b_1-1)-4 b_2 \epsilon$. For cooperation to be sustained, it is required that both the strategies $({\rm TFT};{\rm ALLC})$ and $({\rm ALLC};{\rm TFT})$ are ESS against invasion of mutant $({\rm ALLD};{\rm ALLD})$. So both the aforementioned inequalities have to be satisfied simultaneously for sustenance of cooperation. For the specific choice $N=100$, $b_1=2.0$, $b_2=1.2$, $\bm{q}=\bm{q_{00}}$ and $\epsilon=10^{-3}$ aforementioned inequalities reduce to $0.17<n_{2}<0.83$, which is in good agreement with the findings.

\begin{figure*}[t!]
	\centering
	\includegraphics[scale=0.5]{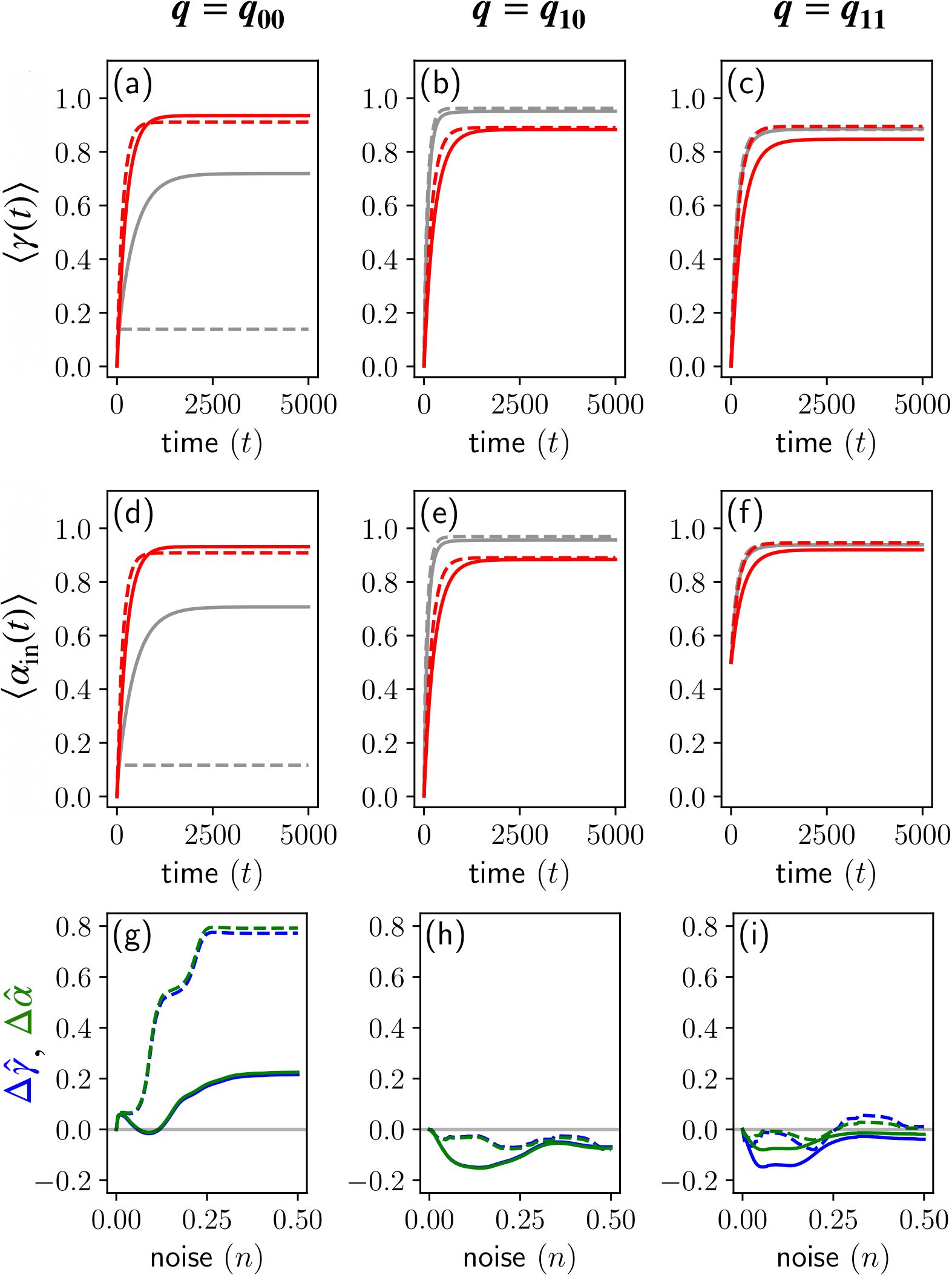}
	\caption{{\it Symmetric noisy channel mediated evolution of cooperation with alternating moves:} The first row (a-c) depicts the evolution of rate of cooperation, $\gamma(t)$, averaged over an ensemble of paths starting from unconditionally defecting resident population. The second row (d-f) exhibits corresponding evolutions of the frequency, $\alpha_{\rm in}$, of beneficial state. The first, the second, and the third columns, respectively, correspond to the outcomes of transition vectors $\bm{q_{00}}$, $\bm{q_{10}}$, and $\bm{q_{11}}$. In these plots, solid and dashed curves represent outcomes for populations with memory-$1$ and memory-$\frac{1}{2}$, respectively. The red and the grey colored curves in the first two rows, respectively, represent outcomes for cases where the information channel is maximally noisy $(n=0.5)$ and minimally noisy $(n=0)$. We observe that for the transition vectors $\bm{q_{00}}$ a noisy channel is beneficial for a population of memory-$\frac{1}{2}$ strategies significantly, contrary to the case of the transition vectors $\bm{q_{10}}$ and $\bm{q_{11}}$. This observation is quantified in the third row (g-i) that depicts the long-run time-averaged enhancement of cooperation, $\Delta\hat{\gamma}$ (blue curves), and of the probability of being in the most beneficial state, $\Delta\hat{\alpha}$ (green curves), for all possible symmetric noisy channels. For illustration purpose, we have fixed $N=100$, $b_{1} = 2.0$, $b_{2}=1.2$, $c=1.0$, $\epsilon=10^{-3}$, and $\beta=10$.}
	\label{afig:symmetricA}
\end{figure*}

\subsection{Minimal representation of the Markov chain with transition matrix ${\sf R}$}\label{aMinimalRep}
In the main text, we have demonstrated that the three primary strategies that govern the asymptotic state of the Markov chain with transition matrix, ${\sf R}$, are $({\rm ALLD};{\rm ALLD})$, $({\rm TFT};{\rm ALLC})$, and $({\rm ALLC};{\rm TFT})$. However, constructing a reduced Markov chain with only these three strategies fails to reproduce the outcomes observed in the main text. By incorporating merely two more strategies, $({\rm TFT};{\rm TFT})$ and $({\rm ALLC};{\rm ALLC})$, into the reduced Markov chain (with transition matrix, say, ${\sf R}_{\rm reduced}$), we achieve outcomes qualitatively almost equivalent to those observed in the Markov chain with transition matrix ${\sf R}$ having all the sixteen strategies (see Fig.~\ref{afig:cr}). Therefore, we conclude that ${\sf R}_{\rm reduced}$ is the appropriate approximation of ${\sf R}$. In summary, to get an understanding of the dynamics, the comprehension of the interplay between these five strategies would come handy.

\begin{figure*}[t!]
	\centering
	\includegraphics[scale=0.4]{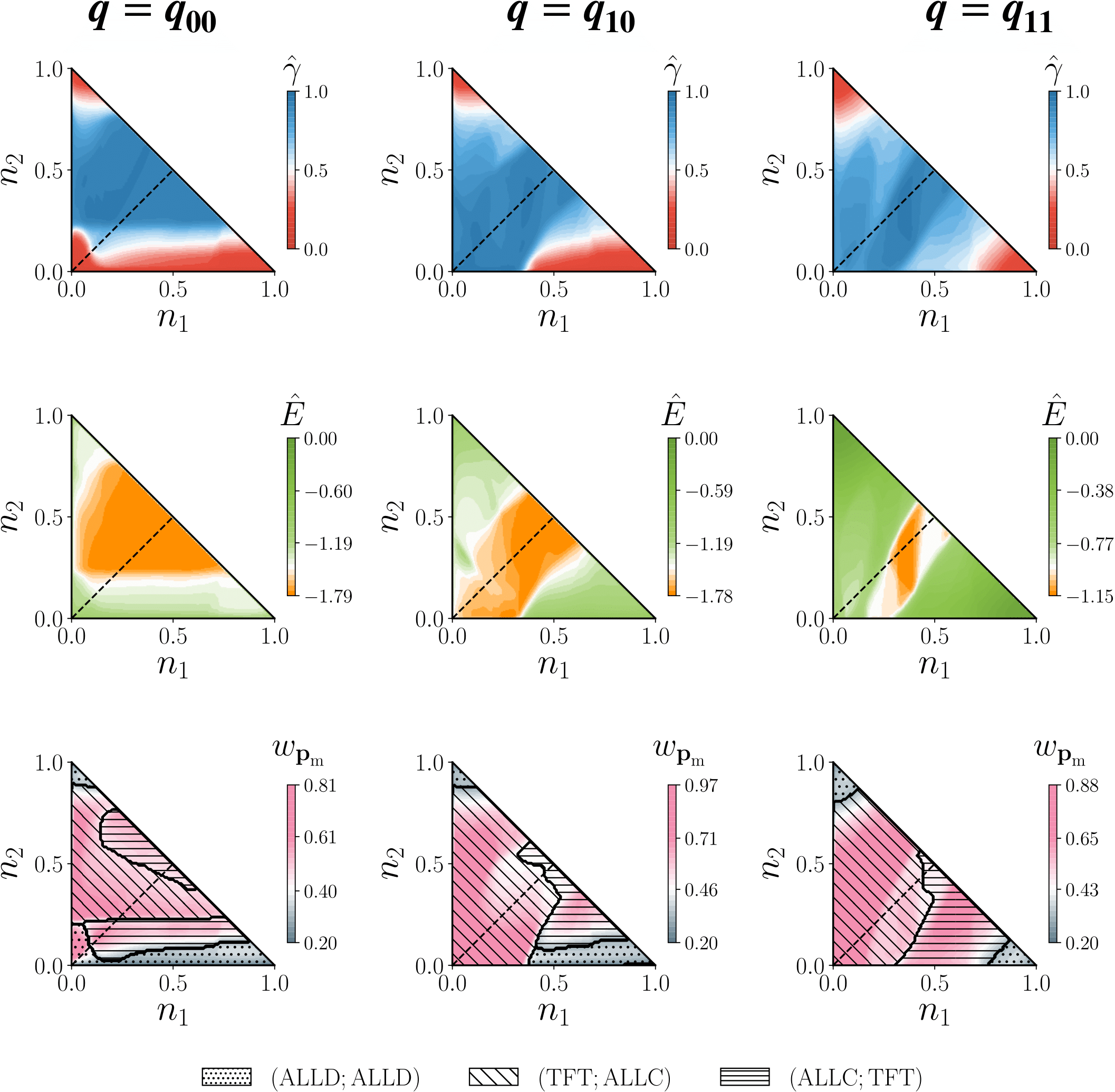}
	\caption{{\it Asymmetric noisy channel mediated evolution of cooperation and efficacy with alternating moves:} The first, second, and third columns represent the outcomes for transition vectors \(\bm{q_{00}}\), \(\bm{q_{10}}\), and \(\bm{q_{11}}\), respectively, for all possible binary asymmetric channels. In the first row (a-c), plots depict the long-run time-averaged cooperation rate, \(\hat{\gamma}\); in the second row (d-f), plots depict the long-run time-averaged efficacy, \(\hat{E}\); and in the third row (g-i), plots depict the maximally recurrent strategies \(\bm{p}_{\rm m}\) and their long run frequencies of recurrence \(w_{\bm{p}_{\rm m}}\) in the population. The dashed line \(n_{1}=n_{2}\) represents subset corresponding to the symmetric channels. For illustration purpose, we have fixed $N=100$, $b_{1} = 2.0$, $b_{2}=1.2$, $c=1.0$, $\epsilon=10^{-3}$, and $\beta=10$.}
	\label{afig:channelA}
\end{figure*}

\subsection{Outcomes corresponding to the case of alternating moves by the players}\label{aResAlt}
In the main text, we have elaborated the result for the case of players with simultaneous moves. Here, we elucidate on the case of players with alternating moves. To begin with, we study the impact of noise by delving into the dynamics of two classes of strategies, namely memory$-\tfrac{1}{2}$ and memory$-1$, across three distinct transition vectors pertinent to the model of the tragedy of the commons.

\subsubsection{Symmetric information channel}
Noise plays a significant role in enhancing the level of cooperation in the $\bm{q_{00}}$-game for both memory-$\tfrac{1}{2}$ and memory-$1$ strategies. However, the enhancement observed in the memory-$\tfrac{1}{2}$ strategy population is notably greater than that in the memory-$1$ strategy population. A common feature of the effects of noise on the $\bm{q_{10}}$-game and $\bm{q_{11}}$-game is that, when memory-1 strategies are employed, noise has a minor detrimental effect on both the stationary cooperation level and the stationary frequency of the beneficial state. This is evident from the third row of Fig.~\ref{afig:symmetricA}, where the changes in equilibrium cooperation rate $\Delta\hat{\gamma}$ (blue solid curve) and the equilibrium frequency of the beneficial state $\Delta\hat{\alpha}$ (green solid curve) fluctuate around zero across all values of $n$. In contrast, for reactive strategies with transition vectors $\bm{q_{10}}$ and $\bm{q_{11}}$, the impact of channel noise is comparatively insignificant relative to memory-$1$ strategies.

\begin{figure*}[t!]
	\centering
	\includegraphics[scale=0.4]{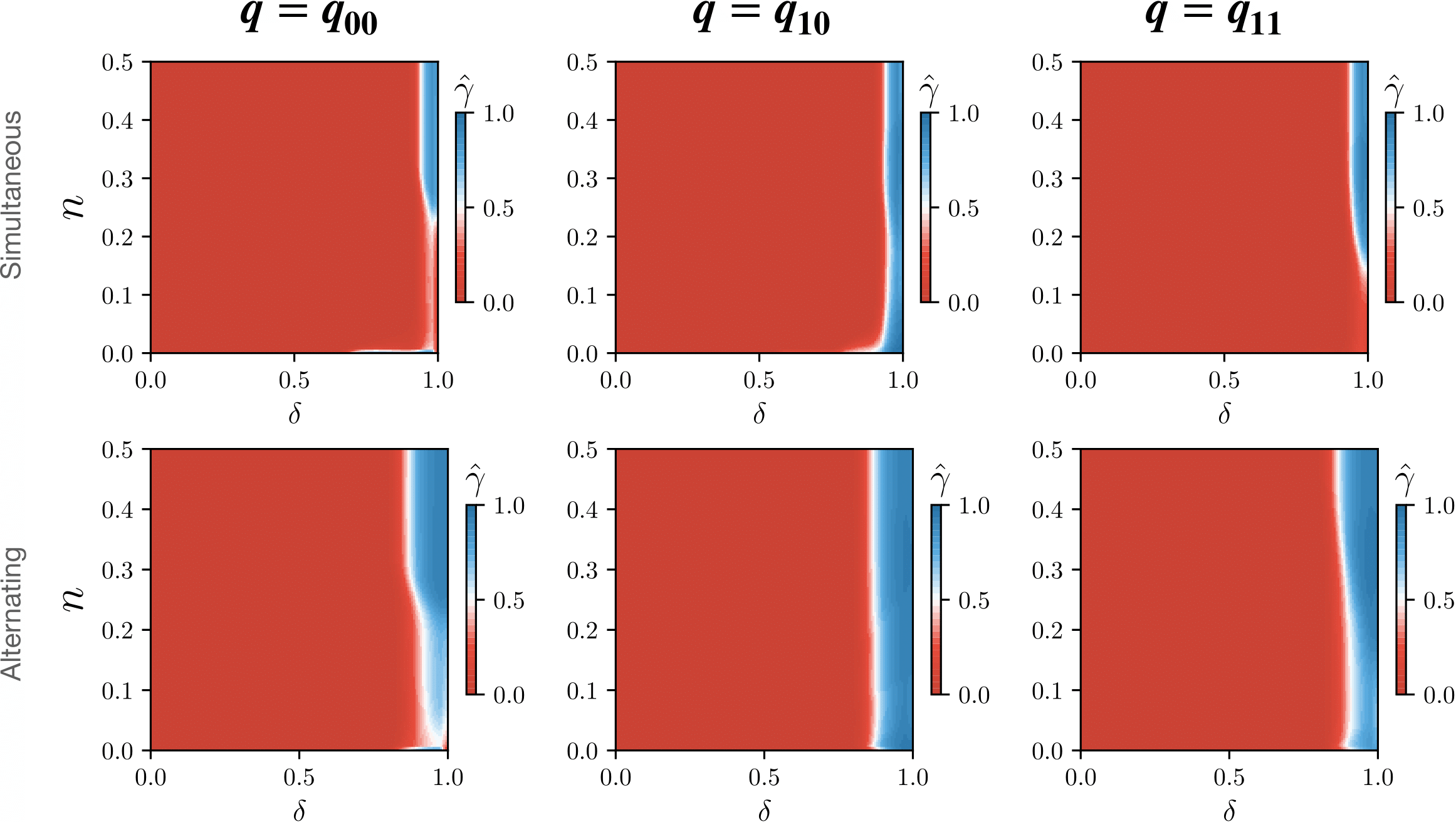}
	\caption{{\it Long-run time averaged cooperation rate for symmetric information channels $n$ with discounted payoffs:} First, second, and third column represent outcomes for transitions vectors $\bm{q}=\bm{q_{00}}$, $\bm{q}=\bm{q_{11}}$, and $\bm{q}=\bm{q_{10}}$ respectively. First and second row depicts the equilibrium cooperation rate $\hat{\gamma}$ for simultaneous and alternating moves respectively. For illustration purpose, we have fixed $N=100$, $b_{1} = 2.0$, $b_{2}=1.2$, $c=1.0$, $\epsilon=10^{-3}$, and $\beta=10$.}
	\label{afig:discountSimAlt}
\end{figure*}
\begin{figure*}[t!]
	\centering
	\includegraphics[scale=0.4]{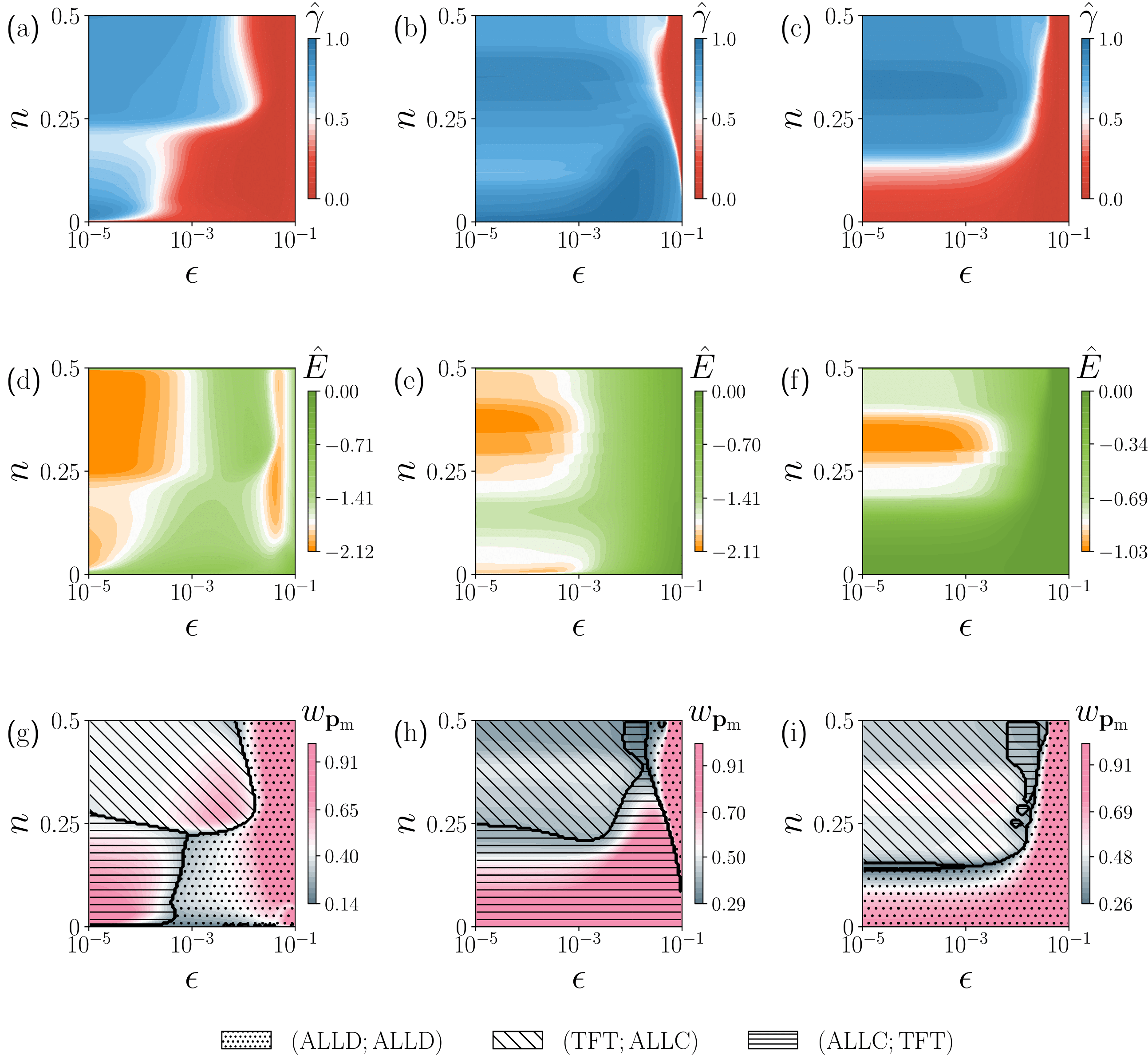}
	\caption{\color{black}{{\it Effect of execution error $\epsilon$ on symmetric noisy channels in simultaneously repeated games:} The first, second, and third columns represent the outcomes for the transition vectors $\bm{q_{00}}$, $\bm{q_{10}}$, and $\bm{q_{11}}$, respectively, across all possible binary symmetric channels (i.e., $n\in[0,1]$) within the considered range of execution errors, $\epsilon \in [10^{-5},\,10^{-1}]$. In the first row (a–c), the plots depict the long-run time-averaged cooperation rate, $\hat{\gamma}$; in the second row (d–f), the long-run time-averaged efficacy, $\hat{E}$; and in the third row (g–i), the maximally recurrent strategies, $\bm{p}_{\rm m}$, along with their long-run recurrence frequencies, $w_{\bm{p}_{\rm m}}$, in the population. For illustration purposes, we have fixed the parameters to $N = 100$, $b_{1} = 2.0$, $b_{2} = 1.2$, $c = 1.0$, and $\beta = 10$.}}
	\label{fig:execError}
\end{figure*}

\subsubsection{Asymmetric information channel}
An observation of the equilibrium cooperation rate $\hat{\gamma}$ and efficacy $\hat{E}$ (see the first and second rows of Fig.~\ref{afig:channelA}, respectively) across all three games reveals a notable characteristic: Whenever non-zero noise induces significant cooperation in the evolutionary system (indicated by blue regions in Fig.~\ref{afig:channelA}a--c), the long-term efficacy in individuals' information processing about the state declines (shown as yellow regions in Fig.~\ref{afig:channelA}d--f), similar to the situation with simultaneous moves discussed in the main text. In essence, evolution tends to favor strategies that receive information about the states with a transmission rate substantially below the channel's capacity. While the inverse relationship between cooperation and efficacy is noteworthy, additional insights can be gleaned from Fig.~\ref{afig:channelA}. Firstly, in the $\bm{q_{00}}$-game, the system does not achieve cooperation without noise, whereas in the $\bm{q_{10}}$-game and $\bm{q_{11}}$-game, significant levels of cooperation are attained even in the absence of noise. Secondly, in all three games, low $n_2$ (indicating less error in perceiving the depleted state) combined with high $n_1$ (indicating high error in perceiving the beneficial state) leads to defection (see the bottom right red corners in the first row of Fig.~\ref{afig:channelA}). Conversely, high $n_2$ combined with low $n_1$ also leads to defection (see the top left red corners in the first row of Fig.~\ref{afig:channelA}). Thirdly, intermediate levels of noise facilitate cooperation within the channel.

\subsubsection{Alternating versus simultaneous games}\label{aDiscussion}
In the main text, we thoroughly examined the scenario of simultaneous moves between players. However, the discussion of alternating moves is elaborated upon only in this Supporting Information text. Here, we briefly summarize the similarities and differences between these two cases. In both scenarios, the set of maximally recurrent strategies remains consistent. Specifically, there are three possible maximally recurrent strategies: $({\rm ALLD};{\rm ALLD})$, $({\rm TFT};{\rm ALLC})$, and $({\rm ALLC};{\rm TFT})$. Regardless of whether the moves are simultaneous or alternating, when non-zero noise induces significant cooperation within the evolutionary system, there is a noted decline in the long-term efficacy of individuals' information processing about the state. In the $\bm{q_{00}}$-game, a population with memory-$\tfrac{1}{2}$ strategies derives significant benefits from non-zero noise, independent of the move scheme employed. In contrast, the $\bm{q_{10}}$-game and $\bm{q_{11}}$-game have different impacts depending on the move scheme. Specifically, in the case of alternating moves, the $\bm{q_{10}}$-game is detrimental, while the $\bm{q_{11}}$-game is beneficial for a population with memory-$\tfrac{1}{2}$ strategies. On the other hand, for simultaneous moves, both games are nearly noise-neutral for a population of memory-$\tfrac{1}{2}$ strategies. Notably, while a memory-$1$ population in the $\bm{q_{00}}$-game remains nearly noise-neutral under simultaneous moves, alternating moves allow for benefits from non-zero noise. We also observe that, in cases with zero and maximally symmetric noise, the $\bm{q_{10}}$-game under alternating moves produces nearly memory-independent outcomes, meaning the results are numerically approximately equal regardless of the class of strategies employed.

\subsection{Comparison of outcomes of simultaneous and alternating moves with Shadow of future}\label{aShadowScheme}
For the case of simultaneous moves, by incorporating the phenomenon of future shadowing, we initially examine the impact of noise intensity and discount factor on evolutionary outcomes. Specifically, we analyze a symmetric information channel denoted by $n$, scrutinizing the influence of $\delta$. Our findings, depicted in first row of Fig.~\ref{afig:discountSimAlt}, reveal that except for exceedingly small number of average rounds, i.e., $\tfrac{1}{1-\delta}\geq 30$ outcomes are similar to the case of $\delta\to1$. However, for $\tfrac{1}{1-\delta}\leq 30$ irrespective of the transition vector and the noise strength $n$, cooperation becomes unachievable by the population. Our findings, for the case of alternating moves, illustrated in second row Fig.~\ref{afig:discountSimAlt}, reveal that except for exceedingly small number of average rounds, i.e., $\tfrac{1}{1-\delta}\geq 10$ outcomes are similar to the case of $\delta\to1$. However, for $\tfrac{1}{1-\delta}\leq 10$ irrespective of the transition vector and the noise strength $n$, cooperation becomes unachievable by the population or become decreased than the case with greater number of rounds. In summary, for both the schemes, roughly speaking, if the discount factor is not close to unity, the long-run time average outcome fails to sustain cooperation.

{\color{black}
\subsection{Effect of execution error}\label{app:ep}
To check how robust our results are to the execution error~$\epsilon$, we first look at its effect on a strategy element $p^{j}_{a\tilde{a}}\in\{0,1\}$. When $\epsilon=0$, the value of $p^{j}_{a\tilde{a}}$ stays the same. As $\epsilon$ increases, this value moves toward $1/2$. At $\epsilon=\tfrac12$, the entry becomes $1/2$ and loses its identity compared to the other elements (see Eq.~\ref{aeq:channelTrans}). For $\epsilon>\tfrac12$, the value shifts toward $1 - p^{j}_{a\tilde{a}}$, which means the strategy effectively flips (e.g., ALLC becomes ALLD and vice versa). Because of this symmetry, the strategy space at $\epsilon=0$ is effectively same as at $\epsilon=1$; hence, it is reasonable to not consider $\epsilon>0.5$ in order to avoid repetitive arguments. Now, in the range $\epsilon<0.5$, we examine how execution error affects the simultaneously repeated stochastic game setting (see Fig.~\ref{fig:execError}). We find that cooperation decreases as $\epsilon$ increases. This is expected, since execution error makes strategies lose their identities and behave more randomly, which breaks the conditions needed for cooperation in our setting.}

\bibliography{mondal_etal.bib}
\end{document}